\documentclass[aps,prl,twocolumn,superscriptaddress,showpacs,showkeys,notitlepage, longbibliography]{revtex4-1}
\usepackage{times,amsmath,amsfonts,amssymb,mathrsfs,graphics,graphicx,color,comment,bm}
\usepackage[next]{inputenc}
\usepackage[dvips]{epsfig}
\usepackage[pdfstartview=FitH]{hyperref}
\graphicspath{{fig/}}
\usepackage{natbib}
\usepackage{pdfsync}
\usepackage{appendix}
\usepackage{textcomp}
\usepackage{epstopdf}
\usepackage{xcolor}
\usepackage{float}
\usepackage{amsmath}
\usepackage{soul}

\usepackage{amssymb}
\usepackage{wasysym}

\begin{document}

\title{Tilted Spirals and Low Temperature Skyrmions in $\text{Cu}_{2}\text{OSeO}_{3}$}

\author{M. Crisanti}
\address{Faculty of Applied Sciences, Delft University of Technology, Mekelweg 15, 2629 JB Delft, The Netherlands}

\author{A. O. Leonov}
\address{Department of Chemistry, Faculty of Science, Hiroshima University, Kagamiyama, Higashi Hiroshima, Hiroshima 739-8526, Japan}
\address{IFW Dresden, Postfach 270016, D-01171 Dresden, Germany}
\address{International Institute for Sustainability with Knotted Chiral Meta Matter, Kagamiyama, Higashi Hiroshima, Hiroshima 739-8526, Japan}

\author{R. Cubitt}
\address{Institut Laue-Langevin, 71 Avenue des Martyrs, CS 20156, 38042 Grenoble, France}

\author{A. Labh}
\address{Faculty of Applied Sciences, Delft University of Technology, Mekelweg 15, 2629 JB Delft, The Netherlands}

\author{H. Wilhelm}
\address{Diamond Light Source Ltd., OX11 0DE Didcot, UK}
\address{Helmholtz-Institute Ulm, Helmholtz-Stra{\ss}e 11, 89081 Ulm, Germany} 

\author{Marcus P. Schmidt}
\address{Max Planck Institute for Chemical Physics of Solids, 01187 Dresden, Germany; }

\author{C. Pappas}
\address{Faculty of Applied Sciences, Delft University of Technology, Mekelweg 15, 2629 JB Delft, The Netherlands}

\begin{abstract}
The bulk helimagnet Cu$_2$OSeO$_3$ represents a unique example in the family of B20 cubic helimagnets exhibiting a tilted spiral and skyrmion phase at low temperatures when the magnetic field is applied along the easy $\langle 001 \rangle$ crystallographic direction. Here we present a systematic study of the stability and ordering of these low temperature magnetic states. We focus our attention on the temperature and field dependencies of the tilted spiral state that we observe persisting up to above  $T=$35~K, i.e. up to higher temperatures than reported so far. We discuss these results in the frame of the phenomenological theory introduced by Dzyaloshinskii in an attempt to reach a quantitative description of the experimental findings. We find that the anisotropy constants, which are the drivers behind the observed behaviour, exhibit a pronounced temperature dependence. This explains the differences in the behaviour observed at high temperatures (above $T=18$~K), where the cubic anisotropy is weak,   and at low temperatures (below $T=18$~K), where a strong cubic anisotropy  induces  an abrupt appearance of the tilted spirals out of the conical state and enhances the stability of skyrmions. 

\end{abstract}

\maketitle
\section{Introduction}

Chiral cubic magnets are at the focus of scientific interest as they are the first systems where chiral magnetic skyrmions have been reported \cite{muhlbauer2009, yu2010, wilhelm2011,seki2012observation}.  These bulk helimagnets host a set of competing interactions: in decreasing order of magnitude, the magnetic exchange interaction, the Dzyaloshinskii-Moriya (DMI) interaction and smaller terms of the magnetic anisotropy. The fine interplay between these energies determines the characteristics of the magnetic field ($H$) - temperature ($T$) phase diagrams of these materials, all showing similar features. Below the ordering temperature  \textit{T}\textsubscript{C} the magnetic ground state involves a long range modulation of the magnetization. For zero applied magnetic field this ground  state consists in a multidomain of spirals with a pitch $L_D$ determined by the ratio between the exchange and DMI energies. In addition to these interactions, for cubic helimagnets, the exchange and cubic anisotropies are responsible for pinning the propagation vector of the spirals along specific directions \cite{bak1980, muhlbauer2009, seki2012observation}.With increasing field,   however, the modulation direction of all spirals aligns  along the direction of the applied field  \cite{plumer1981, bauer2017, maleyev2006, Milde2020}. In this set of materials, skyrmions arranged in hexagonal lattices, are spontaneously stabilised by thermal fluctuations \cite{muhlbauer2009, buhrandt2013} in a small region of the H-T phase diagram, just below \textit{T}\textsubscript{C}, commonly referred to as the skyrmion pocket.  

Given the reduced size of the skyrmion pocket, a great deal of effort has been directed towards the engineering of its size and position to enable future skyrmionic applications. In this context, the application of uniaxial strain \cite{seki2017uniaxial, nakajima2018}, hydrostatic pressure \cite{levatic2016, Crisanti2020}, electric fields \cite{white2014, okamura2016, kruchkov2018, white2018} and chemical doping \cite{stefanvcivc2018, Sukhanov2019} are all viable ways to modify the skyrmion pocket extension and position. 

However, several theoretical models predict the stabilisation of skyrmions in bulk cubic helimagnets over a wide range of temperatures below \textit{T}\textsubscript{C} \cite{bogdanov1989, bogdanov1999, rossler2006, leonovPHD, butenko2010}, and not only in a small region of the phase diagram. 
In particular, low temperature skyrmions (LTS) have been recently observed in Cu$_{2}$OSeO$_{3}$, accompanied by tilted spirals (TS) \cite{qian2018, chacon2018, Halder2018, bannenberg_npj}, when the magnetic field is applied along the easy $\langle 100 \rangle$ crystallographic direction. 
That particular behaviour can be traced back to the quantum-mechanical character of the magnetic building blocks of  Cu$_{2}$OSeO$_{3}$. These are tetrahedra formed by four Cu$^{2+}$ with S=1/2 \cite{janson2014}  leading to a particularly strong interplay between the effective single-ion and exchange anisotropies \cite{qian2018}. Through this mechanism, skyrmions, which are more resilient to anisotropy induced deformations, due to their two-dimensional nature, gain stability \cite{chacon2018, bannenberg_npj}. The tilted spiral state results from the interplay between  cubic and exchange anisotropies and spans a wide range of temperatures, for magnetic fields  between  \textit{H}\textsubscript{C1} and \textit{H}\textsubscript{C2}, which mark the transition between the helical and conical states, and the conical and uniformly magnetised, field polarised, states respectively.

In the following we present a systematic study of the stability and ordering of the low temperature magnetic states in bulk Cu$_{2}$OSeO$_{3}$  that complements previous theoretical and experimental findings \cite{qian2018, chacon2018, Halder2018, bannenberg_npj} as we  focus  on the temperature and field dependencies of the TS state. We find that the reduction of the tilting angle with increasing temperature persists up to much higher temperatures than reported so far, and it is seen as a broadening of the conical scattering.  Furthermore, our results show that the transition from the conical to the tilted spiral state changes  with temperature, from second order above 18~K to first order at lower temperatures.

We discuss these results in the frame of the phenomenological theory introduced by Dzyaloshinskii \cite{Dzyaloshinskii1965a}, which captures the main features of our experimental findings. As compared with previous work we go a step further towards a quantitative comparison between  model and  experiment. This leads  to the conclusion that the anisotropy constants, which are the drivers behind the observed behaviour, exhibit a pronounced temperature dependence explaining the difference between the high temperature (low cubic anisotropy) and low temperature (high cubic anisotropy) behaviour, separated at the temperature of 18~K, which corresponds to a critical point of our model. Our approach provides a strategy for an in-depth and quantitative understanding of chiral magnets in view of tailoring their properties for future applications.

\begin{figure}
    \centering
    \includegraphics[width = 0.45 \textwidth]{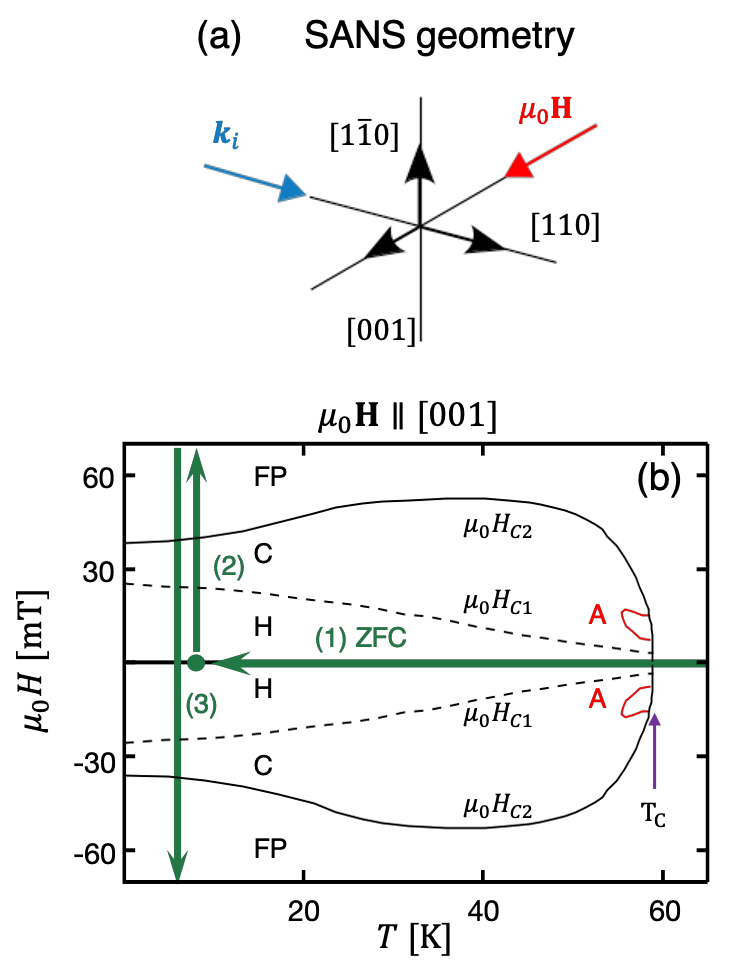}
    \caption{\textbf{(a)}: Schematic representation of the experimental setup, with the $[1\bar{1}0]$ crystallographic axis vertical and perpendicular to the incoming neutron beam propagation vector $\textbf{k}_i \parallel [110]$, and to the  magnetic field $\mu_0 \textbf{H}$, which was applied along $[001]$. 
    \textbf{(b)}: Phase diagram of Cu$_{2}$OSeO$_{3}$ for $\textbf{H} \parallel [001]$ (after \cite{bannenberg_npj,qian2018}) and schematic representation, by the green arrows, of the experimental procedure used for the SANS measurements. In step 1 the sample was brought to each temperature of interest after being  cooled under zero field (ZFC) through $T_C$. Consequently, the magnetic field was  increased up to 74~mT, crossing the critical field lines  $\mu_0 {H}_{C1}$, from the helical (H) to the conical (C) phase, and  $\mu_0 {H}_{C2}$ from the conical to the field polarised (FP) state (step 2). The measurements were then performed by step-wise decreasing the magnetic field to -74~mT (step 3). ). The Figure also depicts in red the boundaries of the A-phase, i.e. the pocket near $T_C$ where the skyrmion lattice phase is stabilised. }
     \label{fig:intro}
\end{figure}

\section{Experimental details}

The small angle neutron scattering measurements were performed at the D33 instrument of the Institut Laue Langevin (ILL), using a wavelength of $\lambda$ = 0.8~nm, with a resolution $\Delta\lambda/\lambda$ of 10~\%. The sample to detector distance was 12.8~m and the sample was the same single crystal of Cu$_{2}$OSeO$_{3}$ used in our previous work \cite{bannenberg_npj,qian2018}. The sample was glued on an Al support with the [1$\bar{1}$0] main crystallographic direction vertical. It  was placed inside an Oxford Instruments 7~T horizontal-field cryomagnet, which was demagnetized at the beginning of the measurements so that the remanent field  was less than 0.1 mT. The cryomagnet was equipped with sapphire windows, each with an opening of $\pm~7^\circ$, 90$^\circ$ apart from each other, which allowed for applying the magnetic field, $\mu_0 \textbf{H}$, either perpendicular or parallel to the incident neutron beam wavevector $\textbf{k}_i$. 

Our SANS measurements were performed in the configuration schematically shown in Figure~\ref{fig:intro}(a), with $\mu_0 \textbf{H} \| [001] $ and $\mu_0 \textbf{H} \perp\textbf{k}_i$. As discussed below, this configuration allows for following the evolution of the helical, conical and tilted spiral Bragg peaks as well as of the skyrmionic scattering when varying the temperature and the magnetic field. In order to determine the scattered intensities associated with these different phases we performed rocking scans by rotating both sample and field around the vertical  [1$\Bar{1}$0] axis in steps of 1 degree, so that $\mu_0 \textbf{H}$ was always parallel to $[001]$. 
Due to the limited opening of the cryomagnet windows these scans covered the angular range of $\pm ~7^\circ$.  At each temperature and magnetic field the  SANS patterns were obtained by summing  the intensities of all rocking scans. For the background correction we used data collected under similar conditions at 10~K and under an applied magnetic field of 2~T.

\begin{figure*}
    \includegraphics[width=0.99\textwidth]{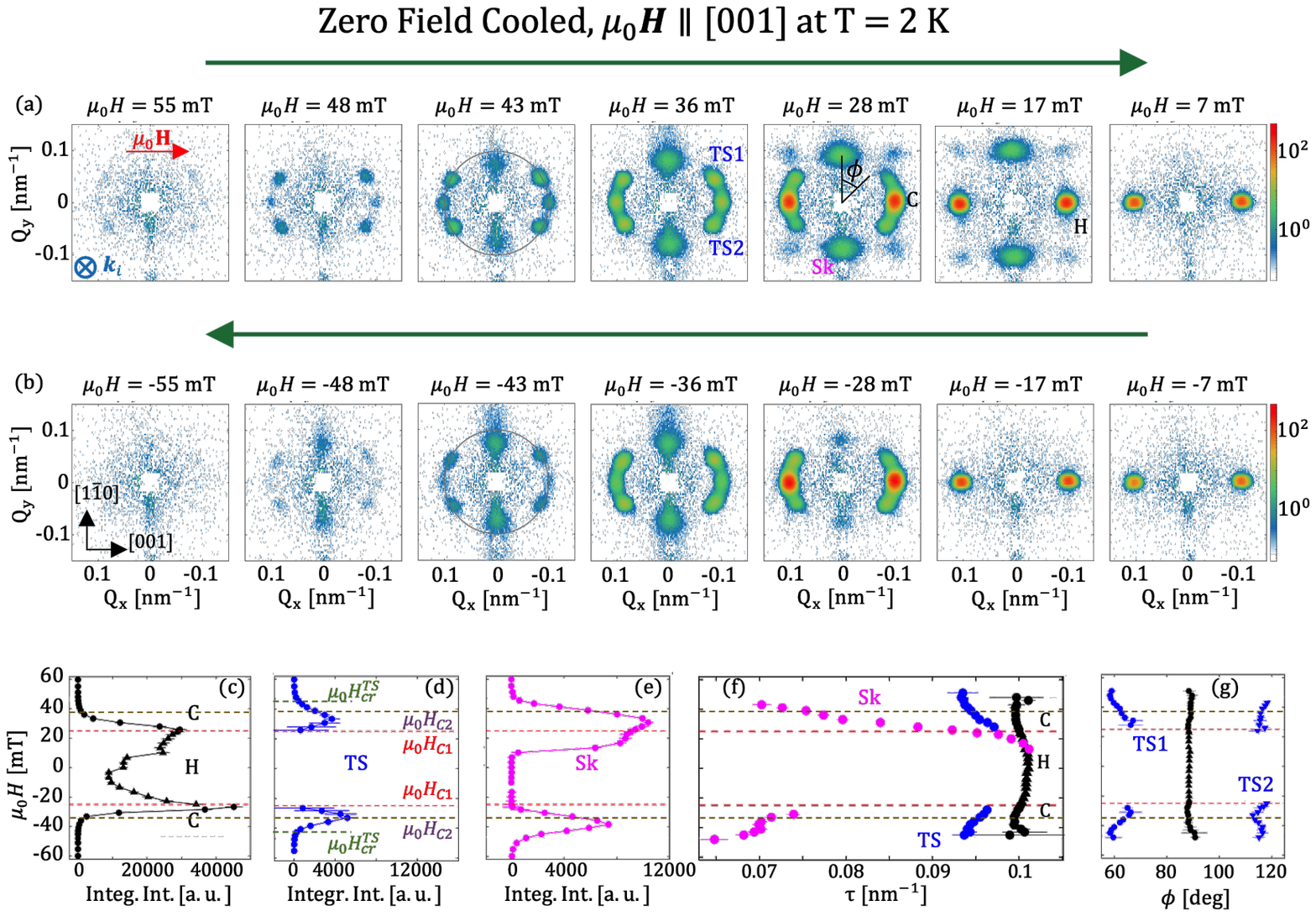}
    \caption{ (a) and (b) show SANS patterns recorded at $T = 2$~K for selected fields using the protocol schematically depicted in Figure~\ref{fig:intro}. The arrows above the SANS patterns indicate the direction of change of the magnetic field. The red arrow in the $\mu_0 H = 57$~mT panel indicates the direction of the applied magnetic field, while $k_i$ indicates the direction of the incoming neutron beam. The field dependence of the integrated intensities of the conical (C), helical (H), TS and skyrmions peaks are shown in panels (c), (d) and (e), respectively. The corresponding values of $\tau$ are provided in (f) and (g) shows the azimuthal position of the conical/helical and TS Bragg spots.In 
    (c-g) the horizontal dashed lines  indicate the critical fields $\mu_0 H_{C1}$ and $\mu_0 H_{C2}$, respectively. The green dashed lines in (d) stand for $\mu_0 H_{cr}^{TS}$. }
    \label{fig:panel2K}
\end{figure*}

\begin{figure*}
    \includegraphics[width=0.99\textwidth]{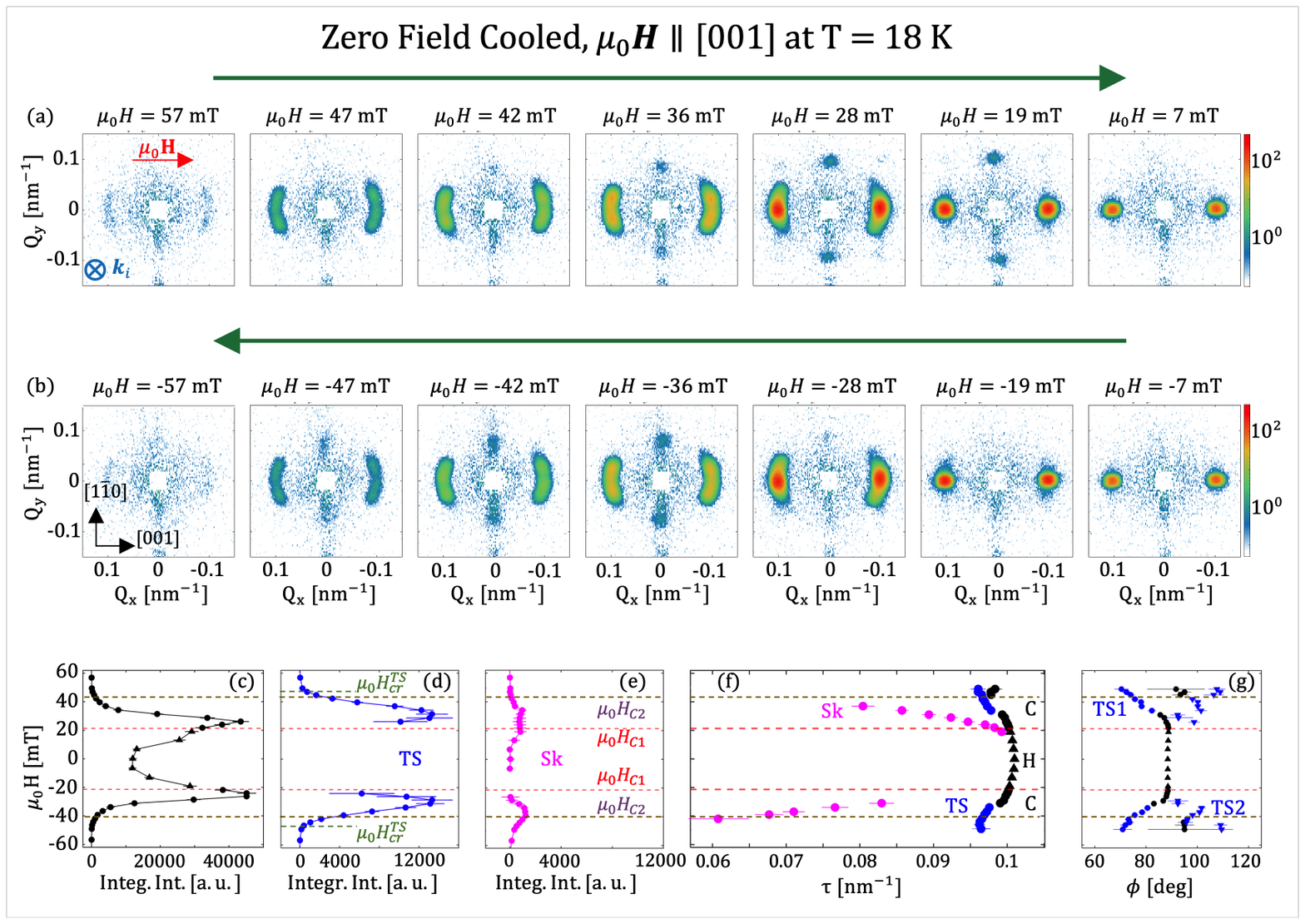}
    \caption{ (a) and (b) show SANS patterns recorded at $T = 18$~K for selected fields using the protocol schematically depicted in Figure~\ref{fig:intro}. The arrows above the SANS patterns indicate the direction of change of the magnetic field. The red arrow in the $\mu_0 H = 57$~mT panel indicates the direction of the applied magnetic field, while $k_i$ the direction of the incoming neutron beam. The field dependency of the integrated intensities of the conical (C), helical (H), TS and skyrmion peaks are shown in panels (c), (d) and (e), respectively. The corresponding values of the  characteristic wavenumbers $\tau$ are provided in (f). (g) shows the azimuthal position of the conical/helical and TS peaks. In  \textbf{(c-g)} the horizontal dashed lines  indicate the critical fields $\mu_0 H_{C1}$ and $\mu_0 H_{C2}$, respectively. The green dashed lines in (d) stand for $\mu_0 H_{cr}^{TS}$.   }
    \label{fig:panel18K}
\end{figure*}

The measurements were performed using the procedure illustrated by the green arrows in Figure~\ref{fig:intro}(b). In the panel are also shown the transition lines from the helical to the conical phase, $\mu_0 H_{C1}$, from the conical to the field polarised phase $\mu_0 H_{C2}$, as well as the A-phase, corresponding to the SKL pocket. We used the zero field cooled, or ZFC, procedure, by bringing the sample to each selected target temperature after cooling it through $T_C \sim 59$~K under zero magnetic field (i.e. under the residual magnetic field of the cryomagnet). Consequently, the magnetic field was increased to 74~mT at 2~K, i.e. well above $\mu_0 H_{C2}$. The measurements were then performed by step-wise decreasing the magnetic field down to -74~mT. The starting and ending values of the magnetic field scans were adjusted to be well above $\mu_0 H_{C2}$ at each target temperature. The magnetic fields have been corrected for demagnetization effects assuming a demagnetization factor of 1/3. \\

\section{Experimental results}

Figure~\ref{fig:panel2K} and \ref{fig:panel18K} 
show SANS patterns collected for $\mu_0 \textbf{H} \parallel [001]$ at $T = 2$ and $18$~K respectively, following the measurement procedure illustrated by Figure~\ref{fig:intro}(b) (for other temperatures the reader is referred to the appendix).  In these Figures, the upper rows (respective panels (a)) show SANS patterns recorded by reducing the magnetic field from well above $\mu_0 H_{C2}$ to zero. The direction of the magnetic field was subsequently reversed and the patterns of the middle rows (respective panels (b)) were recorded for negative fields.
The sequence of the magnetic field change is illustrated by the green arrows above the corresponding SANS patterns and for the sake of comparison, patterns measured for positive magnetic fields are shown above the corresponding pattern for negative fields. This representation highlights the impact of the direction of change of the magnetic field on the observed behaviour. A comparison of Figure~\ref{fig:panel2K} and \ref{fig:panel18K} shows that at $T=2$~K and at intermediate field strengths the SANS patterns strongly depend on the magnetic field sign whereas this effect vanishes at $T=18$~K.

Panels (c), (d) and (e) of Figures~\ref{fig:panel2K} and \ref{fig:panel18K} show the integrated intensities of the conical/helical, tilted spiral and skyrmionic states, deduced from the fitting each scattering spot/peak by a Gaussian function. This fit also provides the characteristic wavenumbers $\tau = 2\pi / L_D$, with $L_D$ the  periodicity of the respective modulations, depicted in (f), as well as the tilting angle of the TS state, reported in (g).

 The moduli of the critical fields, $\mu_0 H_{C2}$ and $\mu_0 H_{cr}^{TS}$, of the transition between the field polarised state on one side and the conical and tilted spiral states on the other side, are estimated from the respective integrated intensities. Within the experimental accuracy these fields correspond to the turning point (ideally the maximum of the second derivative) of the intensity versus magnetic field curves. At 2~K  the values of both  $\mu_0 H_{C2}$ and $\mu_0 H_{cr}^{TS}$ are about 10\%  higher for the positive  than for the negative fields. These differences, which are characteristic of first order transitions decrease with increasing temperature and vanish above $\sim20$~K.

\begin{figure*}
    \includegraphics[width=0.99\textwidth]{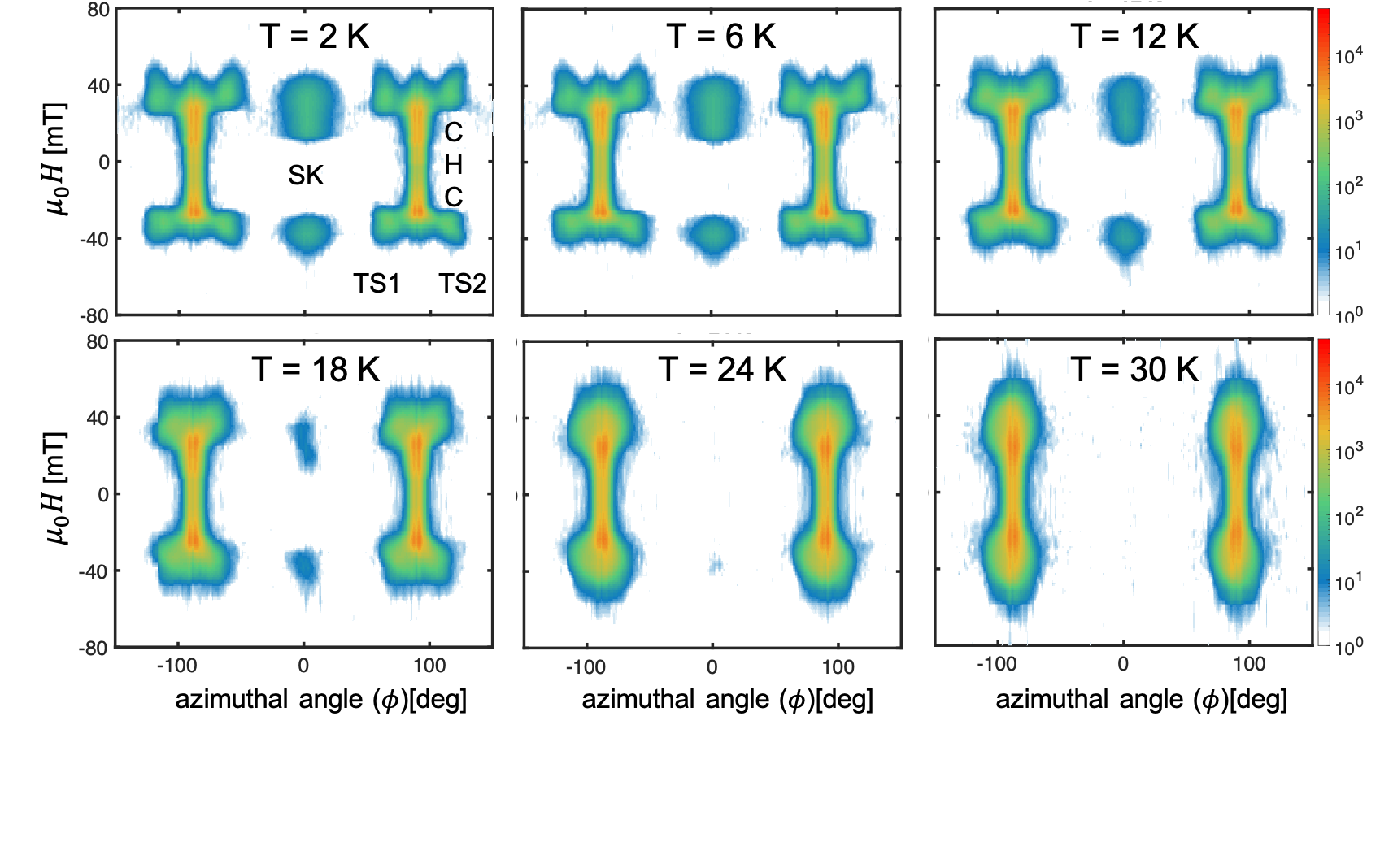}
    \caption{Contour plots of the scattered intensity recorded between  $Q$=0.05 and 0.013~nm$^{-1}$ as a function of  the azimuthal angle $\phi$ and the magnetic field for the temperatures, where measurements were performed.}
    \label{fig:polarPanel}
\end{figure*}

We will in the following look more closely to the evolution of the scattering at $T=2$~K. The patterns collected for high and low fields do not depend on the direction of the magnetic field. At the high field limit, for $\mu_0 H =\pm55$~mT, the modulus of the magnetic field exceeds $\mu_0 H_{C2}$, driving the system to the field polarised state, where the magnetic moments are aligned along the magnetic field and the SANS intensity vanishes. At the low field limit, where $\mu_0 H = \pm7$~mT, the magnetic field modulus is less than $\mu_0 H_{C1}$. Thus the SANS patterns correspond to the helical state, where the spirals are oriented along the equivalent $\langle 001 \rangle$ directions imposed by cubic anisotropy. For the specific sample orientation, only the helical domains aligned along $[001]$ produce scattering, seen as two Bragg peaks in the horizontal direction. In this configuration the Bragg peaks of the helical (e.g. at $\mu_0 H = 0$) and the conical state appear at the same positions on the detector and cannot be distinguished from each other. In the intermediate magnetic field region, with $\mu_0 H_{C1} \lesssim |\mu_0 H| \lesssim \mu_0 H_{C2}$, additional intensity appears besides the conical Bragg peaks, indicating the skyrmionic (Sk) and tilted spiral (TS) scattering. The former is confined in the plane perpendicular to $\mu_0 \textbf{H}$ and appears as two spots along $[1\bar{1}0]$, while the latter consists of four Bragg spots, two around each conical Bragg peak position, from which they are separated by a tilting angle $\alpha = \Delta \phi$. The TS state is almost unaffected by the change of direction of the applied magnetic fields, while both the conical and skyrmionic states are strongly affected by it. This effect can be observed in the integrated intensities reported in panels (c), (d) and (e) for the conical/helical, TS and LTS states.

The integrated intensity of the conical peaks is higher for the negative  magnetic fields than for the corresponding positive fields. This disparity is due to the different profiles of the  rocking curves, that could not be fully probed because of the limited rocking curve measurement range discussed before. Moreover, a small shift of the rocking curves' center was observed between positive and negative fields. In combination with the wide profile of the rocking curves, this could partially explain the observed intensity differences.

As shown in Figure~\ref{fig:panel2K}(d), the tilted spiral scattering appears around $\mu_0H_{C2}$, the intensity increases as the magnetic field is decreased, peaks just before $\mu_0H_{C1}$, and then decreases sharply and disappears at $\mu_0H_{C1}$.  The tilted spiral intensity is recovered when reversing the magnetic field. This scattering thus does not depend on the magnetic history, in contrast, to the intensity of the skyrmionic scattering, shown in Figure~\ref{fig:panel2K}(e). With decreasing the magnetic field, the skyrmionic scattering appears just below  $\mu_0H_{C2}$, reaches its maximum value around $\mu_0H_{C1}$, and persists  even in the helical state, where the intensity decays and disappears just before reaching $\mu_0H = 0$~mT. However, when the field direction is inverted, this scattering emerges only after crossing $-\mu_0H_{C1}$ and almost together with the TS peaks. Like all other contributions, this scattering also disappears as $-\mu_0H_{C2}$ is approached.

The changes in the scattering discussed above also affect the characteristic wavenumbers of the modulations. Near $\mu_0H_{C2}$, as the field decreases following the measurement procedure discussed before, the wavenumber of the conical scattering, 
shown in Figure~\ref{fig:panel2K}(f), first slightly decreases and then increases reaching its maximum value of $10.11(1)\times 10^{-4}$~nm$^{-1}$ at $\mu_0H=0$~mT. A symmetrical behaviour is observed when the magnetic field direction is inverted and the field intensity increased. Similar values and a similar symmetrical behaviour, with respect to the magnetic field direction, are also found for the wavenumber of the tilted spiral scattering. Also this wavenumber increases with decreasing field intensity from $9.34(2)\times  10^{-4}$~nm$^{-1}$ at $\mu_0H_{C2}$ to $ 9.78(1)\times 10^{-4}$~nm$^{-1}$ at $\mu_0H_{C1}$ where it reaches its maximum. 
The wavenumbers of the skyrmionic scattering are significantly lower than those of the other two contributions, as seen e.g. in the SANS patterns for  $\mu_0 H = \pm 43$~mT. At high positive magnetic fields we find  values as low as $7.63(6) \times 10^{-4}$~nm$^{-1}$ which however quickly increase with decreasing  field, tending towards the wavenumbers of the other contributions and reaching a maximum of $9.98(6) \times 10^{-4}$~nm$^{-1}$ at $\mu_0 H =13$~mT. As the field direction is inverted, and the skyrmionic scattering reappears just below $\mu_0H_{C1}$, its wavenumber takes values comparable, and even lower, to those found for high positive fields. 

Finally, Figure~\ref{fig:panel2K}(h) depicts the field dependency of the azimuthal position of the tilted spiral and conical peaks. The latter remain in the same angular position, at $\approx \text{90}^{\circ}$ for all magnetic fields. At the same time, with decreasing the intensity of the magnetic field, the position of the TS peaks oscillates and drifts towards the position of the conical/helical peaks. However, these peaks do not merge with the conical peaks but disappear abruptly at $|\mu_0H_{C1}|$, indicating a discontinuous transition. 

The same measurements have been performed at different temperatures, and the effect of increasing temperature becomes visible at the scattering patterns obtained for $T = 18$~K shown in Figure~\ref{fig:panel18K}. At this temperature the skyrmionic scattering is considerably weaker than at $T=2$~K. At the same time the tilted spiral peaks almost merge with the conical ones and their discrimination becomes difficult, which affects the resulting integrated intensities displayed in Figure~\ref{fig:panel18K}(a) and (b), as well as the deduced azimuthal angles shown in (h). 
At this temperature, the effect of magnetic history on the SANS patterns is almost negligible with the exception of the skyrmionic scattering, which appears over a narrower magnetic field range than at $2$~K, as shown in  Figure~\ref{fig:panel18K}(c), and does not persist into the helical phase when the magnetic field is decreased from $\mu_0H_{C2}$ down to zero.  
On the other hand, the wavenumbers of the different contributions to the scattering and their magnetic field dependence, shown in Figure~\ref{fig:panel18K}(d) and (e), are very similar to those found at $2$~K. 
\begin{figure}
    \centering
    \includegraphics[width=0.9\columnwidth]{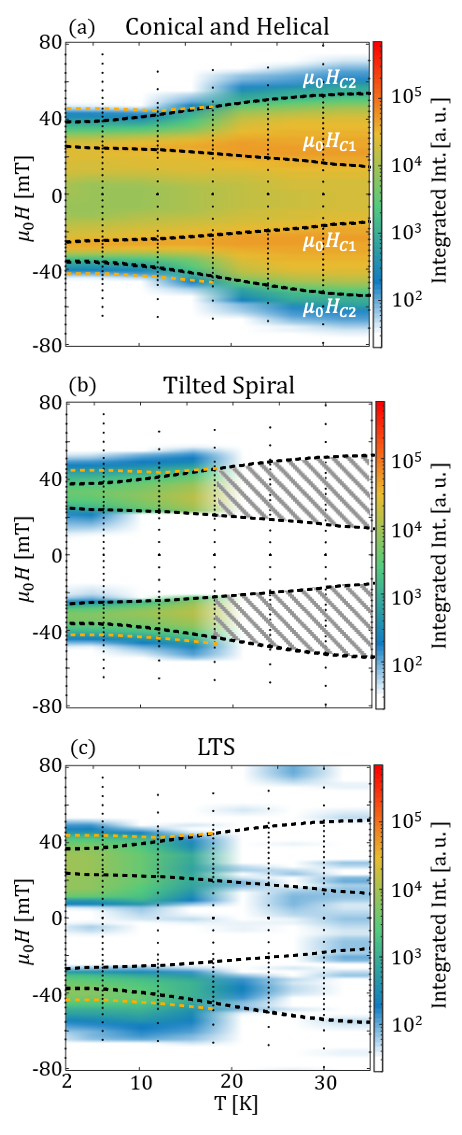}
    \caption{(a-c) Contour plots of the intensities of the conical, tilted spiral and low temperature skyrmionic scattering for $\bf{H} \parallel [001]$. The  black dots indicate the $H-T$ points at which the measurements were performed. The patterned area in panel (b) shows the temperature and field region where the we observe a broadening of the conical peaks and we cannot unambiguosly descriminate the TS from the conical peaks. The black and orange dotted lines indicate  $\mu_0 H_{C2}$ and $\mu_0 H_{cr}$ respectively.}
    \label{fig:phaseDiag}
\end{figure}

The most striking difference between $2$ and $18$~K is in the magnetic field dependence of the azimuthal position of the tilted spiral scattering depicted in Figure~\ref{fig:panel18K}(h). Despite the difficulty in discriminating the peaks, there is a clear indication for a gradual merging of the tilted spiral and conical/helical spots with decreasing $|\mu_0 H|$, in contrast to the abrupt change found at $2$~K. The way the tilted spiral peaks emerge out of the conical/helical scattering as a function of magnetic field and temperature is illustrated in the contour plots of the scattered intensity versus magnetic field and azimuthal angle ($\phi$) shown in Figure \ref{fig:polarPanel}. At $T = 2$~K the tilted spiral scattering  appears in an abrupt, step-wise, manner and is well separated from the conical/helical peaks that are centered at $\phi=0^{\circ}$ and 180$^{\circ}$. As the temperature increases, the angular separation between the conical/helical and the tilted spiral peaks decreases and the transition becomes gradual. Already at $18$~K it is almost impossible to separate the peaks from each other and at higher temperatures the tilted spiral phase appears as a broadening of the helical/conical peaks, an effect that spans the whole conical phase and persists up to above 35~K (see appendix), the highest temperature where our measurements have been performed.    

Figure~\ref{fig:phaseDiag} shows the contour plots of the intensities deduced from our fits: the conical  intensity (along [001]) is shown in Figure~\ref{fig:phaseDiag}(a) and vanishes above  $\mu_0H_{C2}$.   
The TS  intensity shown in  Figure~\ref{fig:phaseDiag}(b) persists slightly above the $\mu_0H_{C2}$ and disappears for field intensities lower than $\mu_0H_{C1}$. The patterned area in this panel identifies the region where it was impossible to distinguish between the conical and TS states. The magnetic field and temperature dependence of the skyrmionic intensity is  shown in Figure~\ref{fig:phaseDiag}(c). This intensity is more extended in the positive fields region, where it is present also below $\mu_0H_{C1}$, compared to the negative fields, where it  appears confined between $-\mu_0H_{C1}$ and $-\mu_0H_{C2}$. This highlights the highly hysteretic behaviour of this scattering.

\section{Phenomenological Model}


In the following we discuss these results in the frame of the phenomenological theory introduced by   Dzyaloshinskii \cite{Dzyaloshinskii1965a}. The magnetic energy density of a bulk non-centrosymmetric ferromagnet with spatially dependent magnetization vector $\mathbf{M}$ can be written as:

\begin{equation}
W_0(\mathbf{m})=A \sum_{i,j}\left(\frac{\partial m_j}{\partial x_i}\right)^2
+D\,w_D(\mathbf{m}) -\mu_0 M_0 \mathbf{m}\cdot\mathbf{H},
\label{density}
\end{equation}
where $A$ and $D$ are the coefficients  of the exchange and Dzyaloshinskii-Moriya interactions respectively; $\mathbf{m}= \mathbf{M}/M_0 = (\sin\theta\cos\psi;\sin\theta\sin\psi;\cos\theta)$ with $M_0 = 111.348$~kA/m the saturation magnetisation at $T=0$~K; $\mathbf{H} = (0,0,H)$  is the  applied magnetic field and $x_i$ are the Cartesian components of the spatial variable.  We will also express our results as a function of the reduced magnetic field  $\mathbf{h}=\mathbf{H}/H_D$ with $\mu_0 H_D = D^2/(A M)$. 

The functional $w_D$ is composed of Lifshitz invariants $\mathcal{L}^{(k)}_{i,j} = m_i \partial m_j/\partial x_k - m_j  \partial m_i/\partial x_k$ that are energy terms involving first derivatives of the magnetization  with respect to the reduced spatial coordinates (normalised to the lattice constant $a$). Consequently, the  sign of $D$  determines the sense of the magnetization rotation. 

$W_0(\mathbf{m})$ includes only basic interactions essential to stabilize skyrmion and helical states and specifies their most general features attributed to all chiral ferromagnets. Our calculations are performed for cubic helimagnets with $w_D=\mathbf{m}\cdot\nabla\times\mathbf{m}$. However, the results are of more general validity and may be applied to magnets with other symmetry classes \cite{bogdanov1989} including different combinations of Lifshitz invariants. 

\begin{figure*}
\includegraphics[width=1.5\columnwidth]{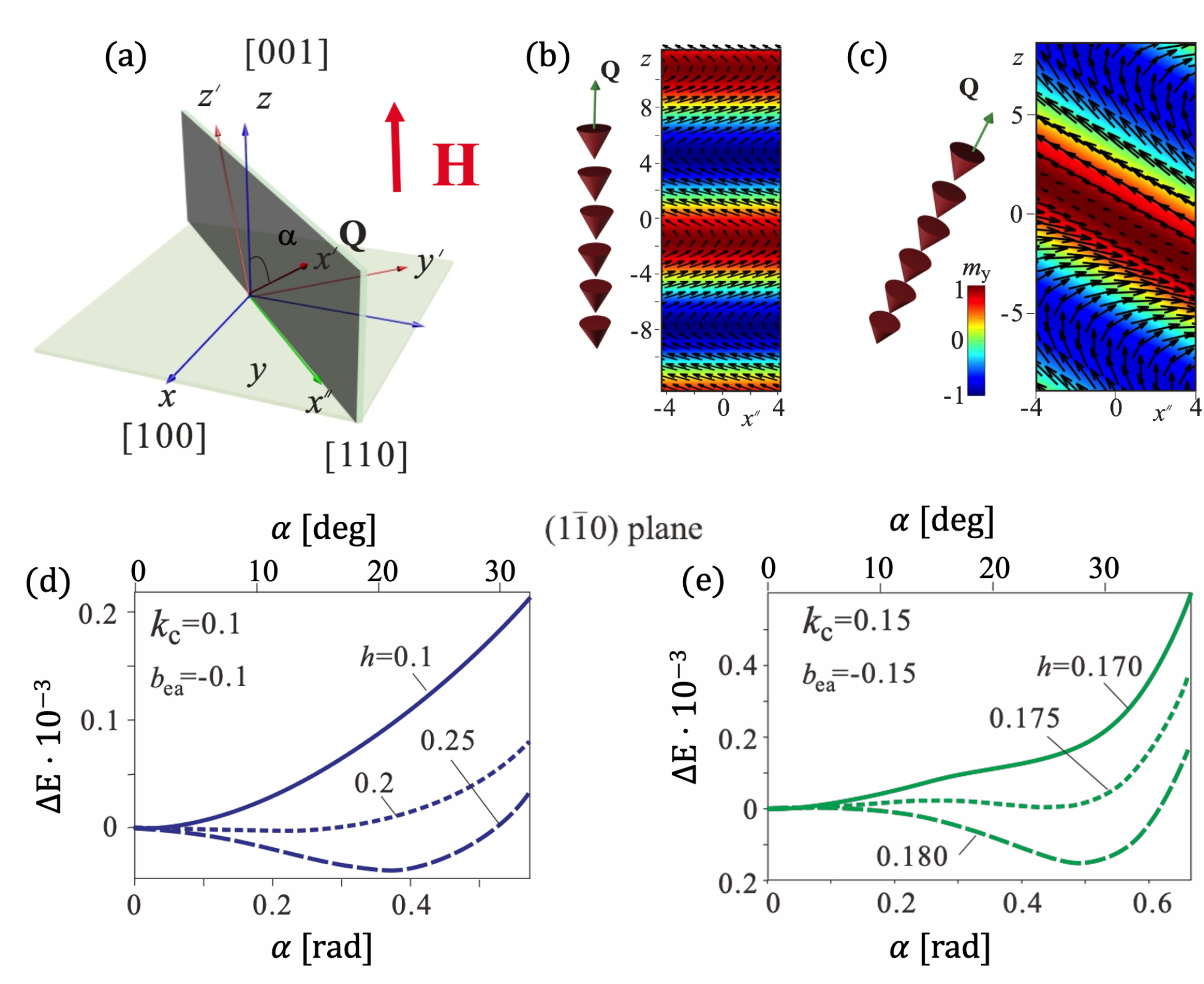}
\caption{ (a) Schematics of a coordinate system used in our numerical search for two-dimensional tilted spiral states. The
energy density is written in the coordinate system ($x'$, $y'$ and $z'$) and is minimized with respect to the angle $\alpha$.  (b), (c) depict sketches of the one-dimensional spiral states - cones (b) and tilted spirals (c) - as well as the corresponding contour plots of the component $m_y$ of the magnetization vector, evaluated in the coordinate system ($x$, $y$, $z$).  (d) and (e) show the energy density plotted as a function of the tilt angle $\alpha$ in the $(1\bar{1}0)$ plane for two representative sets of the anisotropy coefficients ((d) $k_c = 0.1$, $b_{ea} = -0.1$; (e) $k_c = 0.15$, $b_ea=-0.15$) and for several field values highlighting the transition into the tilted spiral state. 
\label{Char01}
}
\end{figure*}

For cubic helimagnets, the isotropic energy density given by eq. \ref{density} is usually supplemented by the exchange and cubic anisotropic contributions \cite{bak1980,janson2014}, 
\begin{equation}
\Phi_a = b_{ea}(T) \sum_{i}(\partial m_i/\partial x_i)^2 +k_c(T) \sum_{i}m_i^4,
\label{anisotropies}
\end{equation}
where $b_{ea}=B_{ea}/A$ and $k_c=K_cA/D^2$ are the reduced exchange and cubic anisotropy constants respectively, which are in general temperature-dependent. 
The constants $k_c$ and $b_{\mathrm{ea}}$ are typically one order of magnitude smaller than the exchange stiffness $A$. 
Neverheless, they play an important role in the stabilisation and control of tilted spiral states.
Furthermore, the cubic anisotropy can  explain the stability of the low temperature skyrmion phase in Cu$_2$OSeO$_3$ \cite{bannenberg_npj}.

For a more direct comparison between experiment and theory we will in the following discuss the results of simulations obtained for $k_c>0$ with easy $\langle001 \rangle$ axes, which corresponds to the case of Cu$_2$OSeO$_3$), and $b_{ea}<0$ with easy $\langle 111 \rangle$ axes. Details on the energy minimisation procedure are provided in the appendix.

In our model the tilted spiral state is found only for $\mathbf{h} \| \langle 001 \rangle$, in agreement with  experiment.
Figure \ref{Char01} (b) and (c) show color plots of the $m_y$ magnetization component in both conical and tilted spirals. 
%
The dependence of the spiral energy on the tilt angle $\alpha$  with the spiral propagation vector  {\bf Q} varying in the (1$\bar 1$0) plane is shown in (d) and (e) for two sets of anisotropy constants and for several field values: (d) $k_c=0.1,\,b_{ea}=-0.1$; (e) $k_c=0.15,\,b_{ea}=-0.15$. In the first case, for $k_c=0.1$ in Figure \ref{Char01}(d), {\bf Q} smoothly moves out of the field direction. Indeed,  the energy difference between the conical state (energy maximum) and an tilted spiral (energy minimum) is very small, and the energy curve exhibits a plateau-like behaviour over the whole range where a tilted spiral can be stabilised. 

For a higher cubic anisotropy, $k_c=0.15$ in (Figure \ref{Char01}(e)), however, there is a small energy barrier between the conical and the tilted spirals (dotted line corresponding to $h=0.175$), which would lead to an abrupt change of the spiral wave vector. Also in this case the energy minimum corresponding to a tilted spiral is quite pronounced  and develops within a rather restricted field range  from  0.17  to 0.18.

In view of these results the behaviour of Cu$_{2}$OSeO$_{3}$, as shown in Figure \ref{fig:polarPanel}, would indicate a temperature dependent $k_c$, the strength of which would decrease with increasing temperature.

\begin{figure*}
\includegraphics[width=1.99\columnwidth]{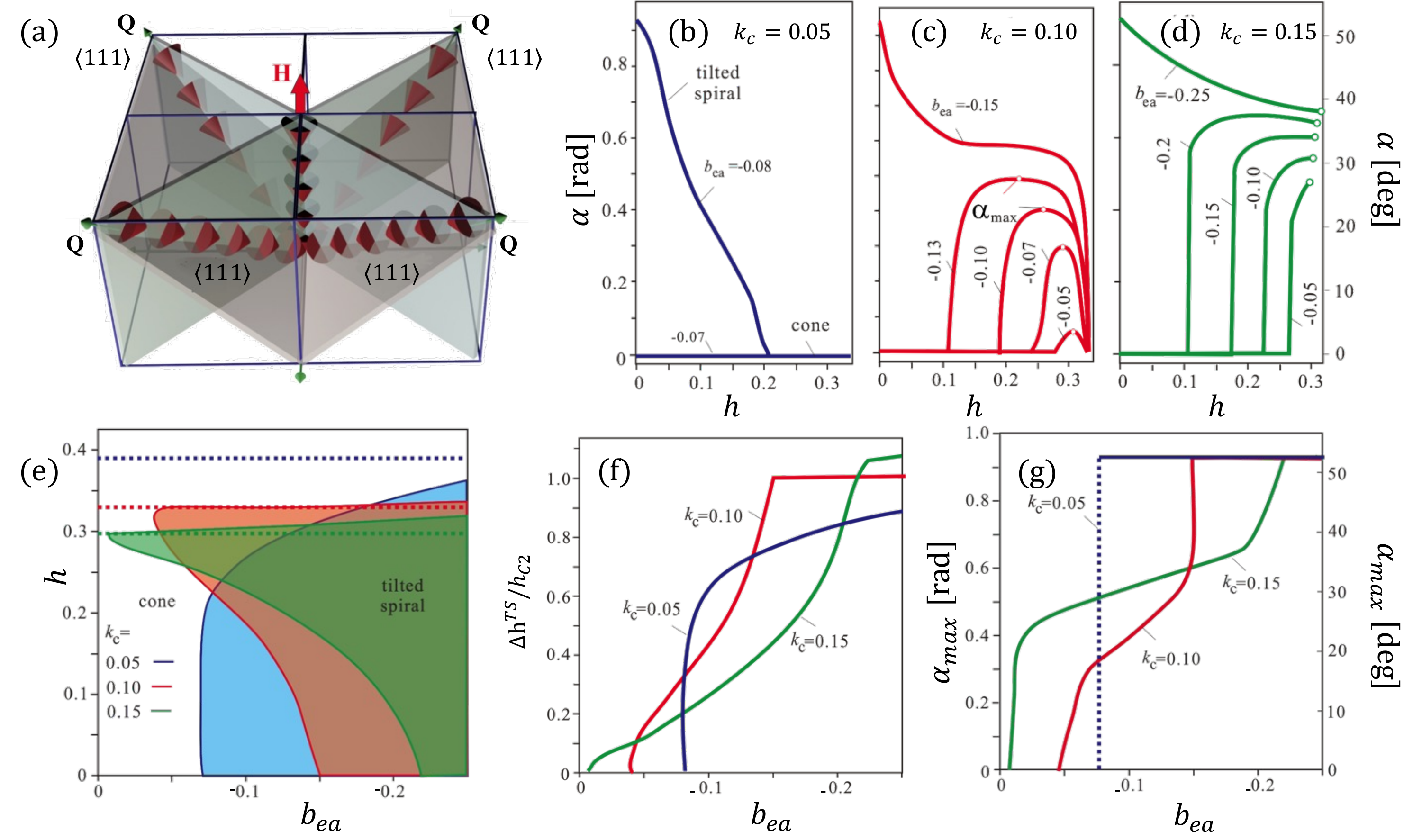}
\caption{Results obtained for $\bf H \parallel [001]$. (a) Schematics of coexisting conical and tilted spiral state consisting of four energetically equivalent domains canting towards the $\langle 111\rangle$ directions. The magnetic field dependence of the tilt angle $\alpha$ is shown in (b), (c) and (d) for cubic anisotropy values of  $k_c$ = 0.05, 0.10 and 0.15 respectively and for selected values of the exchange anisotropy $b_{ea}$. In (c), the tilted spiral almost returns back to the direction of the field, whereas in (d) the spiral is deflected from the field by quite a large angle $\alpha$  and undergoes a first-order phase transition to the field polarised homogeneous state. 
 For the same values of $k_c$  (e) shows a diagram depicting the stability regions of the  the conical and TS states.The dotted lines indicate  $h_{C2}$, which decreases  with increasing $k_c$. The lines  are colored according to the cubic anisotropy value just like the stability regions of the TS phase. The field range, $\Delta h^{TS}$, over which the TS is stabilised, varies strongly with both $k_c$ and $b_{ea}$. The ratio  $\Delta h^{TS} / h_{C2}$ as well as the maximum tilt angle $\alpha_{max}$ are plotted as a function of $b_{ea}$ in (f) and (g), respectively. 
\label{Char02}
}
\end{figure*}

\section{Qualitative model for \NoCaseChange{Cu$_{2}$OSeO$_{3}$} }

For a more quantitative comparison of the model and the experimental findings it is important to  determine the constants in Eqs. (\ref{density}) and (\ref{anisotropies}).
 For the most important Heisenberg exchange and Dzyaloshinskii-Moriya interactions, $A= -k_B J/a$ and $D= k_B D_{DM}/ a^2$, with $a = 0.891113$~nm the lattice constant,  we adopt the values evaluated in Ref. \onlinecite{janson2014}. The variables  $J = -11.19$~K and $D_{DM} = - 2.46$~K were determined using  \textit{ab initio} density functional theory calculations whereas $D_{DM} = - 2.46$~K was evaluated so as to reproduce $L_D = 4 \pi A/D$, the experimentally determined spatial periodicity of the helix.

The important obstacle, however,  lies in the determination of the dimensionless anisotropy constants $k_c$ and $b_{ea}$, as they usually appear together in the magnetization processes and exhibit pronounced temperature dependencies.

In the following, we discuss a strategy that can lead to a complete quantitative model for Cu$_2$OSeO$_3$.
Figure \ref{Char02}(b)  shows that  for a weak cubic anisotropy, e.g. for $k_c = 0.05$, and for sufficiently high values of $|b_{ea}|$  there is an abrupt transtion between the conical phase, where $\alpha=0$ and  the tilted  spiral phase, where spirals orient along $\langle 111 \rangle$.  At a higher value of  $k_c=0.1$ and with increasing $h$ (Figure \ref{Char02}(c)) $\alpha$ first increases reaching a maximal value  of $\alpha_{\rm max}$ before subsequently decreasing back to zero. The value of $\alpha_{\rm max}$  depends on the ratio of the competing fourth-order cubic  and exchange anisotropies, and it is depicted in Figure \ref{Char02}(g). 
As the modulus of exchange anisotropy increases above the critical value of 0.14, the TS state is stabilized even  at zero magnetic field.

By further increasing $k_c$ to 0.15, a qualitatively different behaviour sets-in as shown in Figure \ref{Char02}(d). In this case   with increasing magnetic field  $\mathbf{Q}$ first jumps abruptly to a rather high angle value of  $\alpha$,  about  35 deg for $b_{ea}=-0.15$, and then stays almost unchanged, forming a plateau, up to the critical field of the first-order phase transition to the field polarised state. This behaviour occurs even when the modulus of $b_{ea}$ is smaller than  that of $k_c$, e.g. for $b_{ea}=-0.05$. 
Furthermore, when plotting $\alpha_{max}$ as a function of $b_{ea}$ for  fixed $k_c$ (see Figure \ref{Char02}(g)) we find another extended plateau connecting  the conical phase for low moduli of $b_{ea}$ and the tilted spiral (along  $ \langle 111 \rangle$) for high moduli of $b_{ea}$.

\begin{figure}
\includegraphics[width=0.52\textwidth]{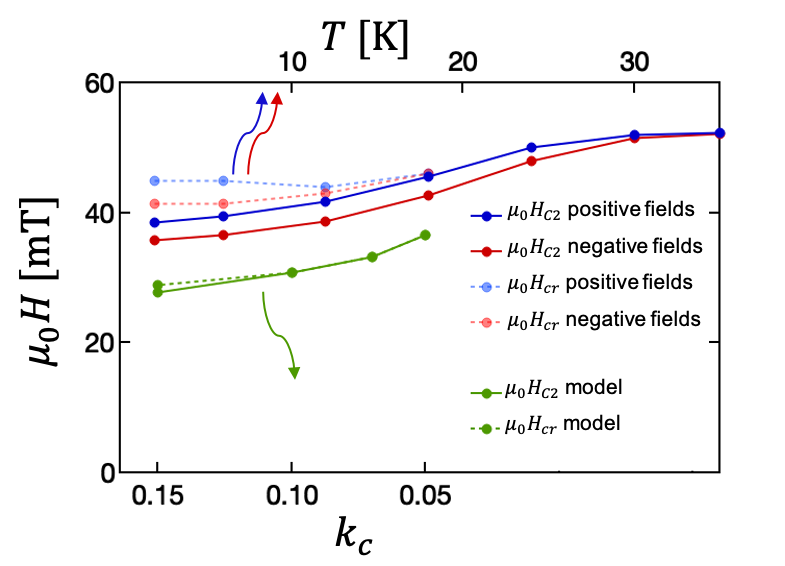}
\caption{ Experimental and theoretical (absolute) values of $\mu_0 H_{C2}$ and $\mu_0 H_{cr}$ plotted as a function of temperature (top axis) and $k_c$ (bottom axis) respectively. The blue and red data represent the experimental values determined for positive (i.e. for decreasing  magnetic field strength) and negative (i.e. for increasing magnetic field strength) fields respectively.  
\label{exp_model}
}
\end{figure}

Figure \ref{Char02}(e) shows the  stability areas of the conical and tilted   spirals for each value of the cubic anisotropy. The dotted lines indicate  $h_{C2}$, the field of the first-order phase transition between the conical and the field polarised state, which decreases substantially with increasing $k_c$.  This figure illustrates the strong dependence of  the field range, $\Delta h^{TS} $, over which the TS is stabilised, on  both $k_c$ and $b_{ea}$. This is also highlighted in Fig. \ref{Char02} ((f), which depicts the ratio  $\Delta h^{TS} / h_{C2}$. Both figures reveal that for $k_c < 0.1$  the upper critical field of the TS phase, $h_{cr}^{TS}$ remains, below $h_{C2}$. For  higher values of $k_c$ however $h_{cr}^{TS}$ increases  and even exceeds   $h_{C2}$. Thus, for the highest moduli of $b_{ea}$, for which the  lower critical field of the TS phase vanishes, the ratio $\Delta h^{TS}/h_{C2} $ may become higher than 1. This  is barely noticeable for $k_c =0.1$ but becomes prominent for $k_c=0.15$. The tilted spiral can therefore transit into the field polarised state at higher fields than the conical state, in agreement with our experimental observations for $T< 18$~K (Fig.  \ref{fig:panel2K}, \ref{fig:panel18K}, \ref{fig:phaseDiag} and \ref{exp_model}). 

A similar change of behaviour with increasing $k_c$ is also found for the maximum tilt angles $\alpha_{max}$. As shown in Fig.  \ref{Char02}(g)  $\alpha_{max}$ strongly increases with increasing $b_{ea}$. However, for $k_c = 0.15$ this increase levels off around  $\alpha_{max} \sim 30$~deg, an effect that becomes less and less visible with increasing $k_c$ and disappears at a critical value of  $k_{c, cr} \sim 0.06$ (see Fig. S5). For $k_c<k_{c, cr}$ one observes just a jump between the two angles that correspond to the $\langle 001 \rangle$ and $ \langle 111 \rangle$ crystallographic axes.

These results lead to the conclusion that $k_{c, cr}$  separates a regime of weak cubic anisotropy  (for $k_c < k_{c, cr} $), where tilted spirals are experimentally barely distinguishable from the cones, from another one of strong cubic anisotropy (for $k_c >  k_{c, cr})$ characterised by well defined  tilted spirals, which appear abruptly out of the conical state and  persist even above $h_{C2}$. We can thus  explain the observed change of behaviour around 18~K by the change from weak to strong cubic anisotropy 
regimes. As the nucleation of skyrmions is assumed  to be related to the stability of tilted spirals \cite{chacon2018, bannenberg_npj, leonov2022prr}, this change of behaviour around 18~K would explain the increased stability of skyrmions at low temperatures.

Further comparison between  model and   experiment  at low temperatures  leads to the following conclusions:\\ 
 (i) A value of $\alpha_{max} \approx 34$~deg, as observed experimentally at 2~K, cannot be reached for $k_c=0.1$, because in that case  with increasing $|b_{ea}|$  the spirals would jump towards the $\langle 111 \rangle$ directions before reaching this angle.   On the other hand, for 
$k_c=0.15$ this angle is readily obtained and with a weak, plateau-like field dependence for $b_{ea} = -0.15$. \\
(ii) At low temperatures, our experimental results show that the TS phase sets-in over a magnetic field interval $\Delta h^{TS} / h_{C2}\approx 0.59$  which, as shown in Fig.  \ref{Char02}(f), leads to $b_{ea}\approx -0.18$,  which is consistent with the estimated value for Zn doped Cu$_2$OSeO$_3$ \cite{Moody2021}. \\

(iii) The relative difference $\delta = \left(h_{cr}^{TS}-h_{C2}\right)/h_{C2}$  is an additional parameter that can be compared with experiment. However, as mentioned above, the experimental values of both critical fields, $\mu_0 H_{C2}$ and $\mu_0 H_{cr}^{TS}$  depend strongly on the magnetic history. At $T=2$~K, we find $\delta \sim$ 22\% and $\sim$ 25\% for  positive and negative fields respectively, a value substantially higher than $\sim$ 2\% derived from our numerical simulations. Furthermore, the difference between $\mu_0 H_{C2}$ and $\mu_0 H_{cr}^{TS}$  persists even at $T=18$~K (see Fig. \ref{fig:panel18K} and \ref{exp_model}), where we find $\delta \sim 10$ \% and $\sim 17$ \% for positive and negative fields, respectively. This hysteretic behaviour reflects the presence of potential barriers and the first-order character of the phase transitions at $h_{C2}$ and $h_{cr}^{TS}$, respectively, and could be at the origin of the discrepancies in the values of $\delta$ between model and experiment.  \\

Our theoretical results for $k_c=0.15$ explain  qualitatively  the following  experimental observations at $T=$ 2~K: \\
(i) The first-order phase transitions to the field polarised, state. The first order nature of these transitions is confirmed by the coexistence of conical and tilted spiral states and the  asymmetry in the moduli of $\mu_0 H_{C2}$ and $\mu_0 H_{cr}^{LS}$ between positive and negative fields. \\
(ii) The abrupt appearance of the tilted spiral state. \\
(iii) The values of the angles and the magnetic field intervals over which the  spiral canting sets-in. 
We note that the value of $k_c=0.15$ exceeds by a factor of two the value of $k_c=0.07$ estimated by \cite{Halder2018} from a comparison between experiment and a model similar to ours but which included dipole interactions and discarded exchange anisotropy. 

With increasing temperature towards 18~K, $k_c$ decreases and the first-order phase transition between the tilted spiral and the field polarised state becomes less obvious because with increasing field the spirals smoothly slant away from the cones  and almost return back to them close to $h_{C2}$. Such a behaviour would correspond to  $k_c=0.1$. As mentioned above this decrease of $k_c$ is consistent with the increase of the experimentally observed values of $\mu_0 H_{C2}$.  The ratio between the $\mu_0 H_{C2}$ values at $T=$18~K and 2~K is about 1.3 and 1.2 for positive and negative fields 
respectively. Thus, if $k_c=0.15$ for $T=2$~K, the corresponding value at $T=$18~K would be equal to $k_{c, cr}=0.06$.  The behaviour experimentally observed for $T>18$~K would therefore correspond to  $k_c < k_{c,cr}$. However, as in the latter case  the TS state does occur at zero field, we deduce that $b_{ea}>-0.08$. 

Our model thus captures the main features of the low temperature behaviour of Cu$_2$OSeO$_3$ . This is illustrated by Fig. \ref{exp_model}, which depicts the (absolute) values of $\mu_0 H_{C2}$ and $\mu_0 H_{cr}$, plotted as a function of temperature and $k_c$ for the  experimental and theoretical values respectively. 

A subsequent endeavor would be to construct a fully quantitative model for a bulk helimagnet Cu$_2$OSeO$_3$ by taking into account experimental data for other field directions  and possibly including dipole-dipole interactions. These have   been considered by \cite{chacon2018, Halder2018} as demagnetising corrections are important in Cu$_2$OSeO$_3$. On the other hand, the phenomena we are discussing here have been reported consistently in the literature for different samples and with different shapes \cite{chacon2018, Halder2018, qian2018, bannenberg_npj}. In addition, the temperature dependence of the magnetization below 30 K is very weak \cite{qian2018, zivkovic2012} in contrast to the observed change in behavior in the same temperature range. We have therefore disregarded these interactions in our model and used this conventional approximation to accelerate the minimization procedure for all states considered here.   

\section{Conclusions}

Our in depth  investigation by small angle neutron scattering and numerical simulations of Cu$_2$OSeO$_3$ reveals an unconventional temperature dependence of the magnetic phase diagram when the magnetic field is applied along $\langle 001 \rangle$. At low temperatures, e.g. $T$ = 2~K, the tilted spiral scattering  appears in an abrupt, step-wise, manner and is well separated from the conical peaks. With increasing temperature however, the angular separation between the conical and the tilted spiral peaks decreases and at $18$~K the transition becomes gradual. Already at $18$~K it is almost impossible to separate the tilted spiral from the conical peaks and at higher temperatures the tilted spiral phase appears as a broadening of the helical/conical peaks, an effect that spans the whole conical phase and persists up to  35~K and possibly above. At all temperatures the tilted spiral phase  appears between $\mu_0H_{C2}$ and $\mu_0H_{C1}$in contrast to the hysteretic behaviour of the  skyrmionic scattering, which appears at different regions of the $H-T$ phase diagram  depending on  magnetic history. 

Our theoretical model based on the theory introduced by Dzyaloshinskii captures the main features of the experimental findings, which can be accounted for by the interplay between exchange and cubic anisotropy. In this work, however, we go a step further towards a quantitative comparison between model and experiment and discuss a strategy that leads to an estimate of the  anisotropy constants. These are the drivers behind the observed behaviour, and their determination is not as trivial as for the interaction constants, that have been obtained either by DFT or experimentally. The comparison between  model and  experiment leads to the conclusion that the anisotropy constants change significantly with temperature and that this temperature dependence explains  the change of behaviour observed around $T \sim 18$~K, a temperature where a critical value of cubic anisotropy, $k_{c,cr} \sim 0.6$,  separates two regimes. The high temperatures one ($T>18$~K) is characterised by weak cubic anisotropy ($k_c < k_{c,cr}$) and encompasses a smooth transition between cones and tilted spirals as well as very weak skyrmionic correlations. In contrast, in the low temperature regime ($T>18$~K) a strong cubic anisotropy ($k_c > k_{c,cr}$) induces an abrupt appearance of the  tilted spirals and enhances the stability of skyrmions. 

The approach we have adopted allows a quantitative comparison between experiment and theory and provides a strategy for an in-depth understanding of chiral magnets in view of tailoring their properties for future applications. 
\\ \\ \\ 
{\bf ACKNOWLEDGEMENTS}
The authors are grateful to  Maxim Mostovoy for useful discussions and to the technical staf of ILL for their support during the experiment. 
A.O.L. thanks Ulrike Nitzsche for technical assistance and acknowledges JSPS Grant-in-Aid (C) No. 21K03406.
M.C, A. L. and C. P acknowledge financial support from 
the Vrije FOM-programma ''Skyrmionics'' (Nr. 838).
\\ \\  \\ 
{\bf DATA AVAILABILITY}
 The raw experimental data is available at doi.org/10.5291/ILL-DATA.5-42-492. Additional data related to this paper may be requested from  M.C. (m.crisanti@tudelft.nl), C.P. (c.pappas@tudelft.nl) and A.O.L. (leonov@hiroshima-u.ac.jp). 
\\ \\  \\ 


\begin{thebibliography}{35}%
\makeatletter
\providecommand \@ifxundefined [1]{%
 \@ifx{#1\undefined}
}%
\providecommand \@ifnum [1]{%
 \ifnum #1\expandafter \@firstoftwo
 \else \expandafter \@secondoftwo
 \fi
}%
\providecommand \@ifx [1]{%
 \ifx #1\expandafter \@firstoftwo
 \else \expandafter \@secondoftwo
 \fi
}%
\providecommand \natexlab [1]{#1}%
\providecommand \enquote  [1]{``#1''}%
\providecommand \bibnamefont  [1]{#1}%
\providecommand \bibfnamefont [1]{#1}%
\providecommand \citenamefont [1]{#1}%
\providecommand \href@noop [0]{\@secondoftwo}%
\providecommand \href [0]{\begingroup \@sanitize@url \@href}%
\providecommand \@href[1]{\@@startlink{#1}\@@href}%
\providecommand \@@href[1]{\endgroup#1\@@endlink}%
\providecommand \@sanitize@url [0]{\catcode `\\12\catcode `\$12\catcode
  `\&12\catcode `\#12\catcode `\^12\catcode `\_12\catcode `\%12\relax}%
\providecommand \@@startlink[1]{}%
\providecommand \@@endlink[0]{}%
\providecommand \url  [0]{\begingroup\@sanitize@url \@url }%
\providecommand \@url [1]{\endgroup\@href {#1}{\urlprefix }}%
\providecommand \urlprefix  [0]{URL }%
\providecommand \Eprint [0]{\href }%
\providecommand \doibase [0]{http://dx.doi.org/}%
\providecommand \selectlanguage [0]{\@gobble}%
\providecommand \bibinfo  [0]{\@secondoftwo}%
\providecommand \bibfield  [0]{\@secondoftwo}%
\providecommand \translation [1]{[#1]}%
\providecommand \BibitemOpen [0]{}%
\providecommand \bibitemStop [0]{}%
\providecommand \bibitemNoStop [0]{.\EOS\space}%
\providecommand \EOS [0]{\spacefactor3000\relax}%
\providecommand \BibitemShut  [1]{\csname bibitem#1\endcsname}%
\let\auto@bib@innerbib\@empty
\bibitem [{\citenamefont {M{\"u}hlbauer}\ \emph {et~al.}(2009)\citenamefont
  {M{\"u}hlbauer}, \citenamefont {Binz}, \citenamefont {Jonietz}, \citenamefont
  {Pfleiderer}, \citenamefont {Rosch}, \citenamefont {Neubauer}, \citenamefont
  {Georgii},\ and\ \citenamefont {B{\"o}ni}}]{muhlbauer2009}%
  \BibitemOpen
  \bibfield  {author} {\bibinfo {author} {\bibfnamefont {S.}~\bibnamefont
  {M{\"u}hlbauer}}, \bibinfo {author} {\bibfnamefont {B.}~\bibnamefont {Binz}},
  \bibinfo {author} {\bibfnamefont {F.}~\bibnamefont {Jonietz}}, \bibinfo
  {author} {\bibfnamefont {C.}~\bibnamefont {Pfleiderer}}, \bibinfo {author}
  {\bibfnamefont {A.}~\bibnamefont {Rosch}}, \bibinfo {author} {\bibfnamefont
  {A.}~\bibnamefont {Neubauer}}, \bibinfo {author} {\bibfnamefont
  {R.}~\bibnamefont {Georgii}}, \ and\ \bibinfo {author} {\bibfnamefont
  {P.}~\bibnamefont {B{\"o}ni}},\ }\bibfield  {title} {\enquote {\bibinfo
  {title} {{Skyrmion Lattice in a Chiral Magnet}},}\ }\href@noop {} {\bibfield
  {journal} {\bibinfo  {journal} {Science}\ }\textbf {\bibinfo {volume}
  {323}},\ \bibinfo {pages} {915--919} (\bibinfo {year} {2009})}\BibitemShut
  {NoStop}%
\bibitem [{\citenamefont {Yu}\ \emph {et~al.}(2010)\citenamefont {Yu},
  \citenamefont {Onose}, \citenamefont {Kanazawa}, \citenamefont {Park},
  \citenamefont {Han}, \citenamefont {Matsui}, \citenamefont {Nagaosa},\ and\
  \citenamefont {Tokura}}]{yu2010}%
  \BibitemOpen
  \bibfield  {author} {\bibinfo {author} {\bibfnamefont {X.~Z.}\ \bibnamefont
  {Yu}}, \bibinfo {author} {\bibfnamefont {Y.}~\bibnamefont {Onose}}, \bibinfo
  {author} {\bibfnamefont {N.}~\bibnamefont {Kanazawa}}, \bibinfo {author}
  {\bibfnamefont {J.~H.}\ \bibnamefont {Park}}, \bibinfo {author}
  {\bibfnamefont {J.~H.}\ \bibnamefont {Han}}, \bibinfo {author} {\bibfnamefont
  {Y.}~\bibnamefont {Matsui}}, \bibinfo {author} {\bibfnamefont
  {N.}~\bibnamefont {Nagaosa}}, \ and\ \bibinfo {author} {\bibfnamefont
  {Y.}~\bibnamefont {Tokura}},\ }\bibfield  {title} {\enquote {\bibinfo {title}
  {{Real-space observation of a two-dimensional skyrmion crystal}},}\
  }\href@noop {} {\bibfield  {journal} {\bibinfo  {journal} {Nature}\ }\textbf
  {\bibinfo {volume} {465}},\ \bibinfo {pages} {901--904} (\bibinfo {year}
  {2010})}\BibitemShut {NoStop}%
\bibitem [{\citenamefont {Wilhelm}\ \emph {et~al.}(2011)\citenamefont
  {Wilhelm}, \citenamefont {Baenitz}, \citenamefont {Schmidt}, \citenamefont
  {R{\"o}{\ss}ler}, \citenamefont {Leonov},\ and\ \citenamefont
  {Bogdanov}}]{wilhelm2011}%
  \BibitemOpen
  \bibfield  {author} {\bibinfo {author} {\bibfnamefont {H.}~\bibnamefont
  {Wilhelm}}, \bibinfo {author} {\bibfnamefont {M.}~\bibnamefont {Baenitz}},
  \bibinfo {author} {\bibfnamefont {M.}~\bibnamefont {Schmidt}}, \bibinfo
  {author} {\bibfnamefont {U.~K.}\ \bibnamefont {R{\"o}{\ss}ler}}, \bibinfo
  {author} {\bibfnamefont {A.~A.}\ \bibnamefont {Leonov}}, \ and\ \bibinfo
  {author} {\bibfnamefont {A.~N.}\ \bibnamefont {Bogdanov}},\ }\bibfield
  {title} {\enquote {\bibinfo {title} {{Precursor Phenomena at the Magnetic
  Ordering of the Cubic Helimagnet FeGe}},}\ }\href@noop {} {\bibfield
  {journal} {\bibinfo  {journal} {Phys. Rev. Lett.}\ }\textbf {\bibinfo
  {volume} {107}},\ \bibinfo {pages} {127203} (\bibinfo {year}
  {2011})}\BibitemShut {NoStop}%
\bibitem [{\citenamefont {Seki}\ \emph {et~al.}(2012)\citenamefont {Seki},
  \citenamefont {Yu}, \citenamefont {Ishiwata},\ and\ \citenamefont
  {Tokura}}]{seki2012observation}%
  \BibitemOpen
  \bibfield  {author} {\bibinfo {author} {\bibfnamefont {S.}~\bibnamefont
  {Seki}}, \bibinfo {author} {\bibfnamefont {X.~Z.}\ \bibnamefont {Yu}},
  \bibinfo {author} {\bibfnamefont {S.}~\bibnamefont {Ishiwata}}, \ and\
  \bibinfo {author} {\bibfnamefont {Y.}~\bibnamefont {Tokura}},\ }\bibfield
  {title} {\enquote {\bibinfo {title} {{Observation of Skyrmions in a
  Multiferroic Material}},}\ }\href@noop {} {\bibfield  {journal} {\bibinfo
  {journal} {Science}\ }\textbf {\bibinfo {volume} {336}},\ \bibinfo {pages}
  {198--201} (\bibinfo {year} {2012})}\BibitemShut {NoStop}%
\bibitem [{\citenamefont {Bak}\ and\ \citenamefont {Jensen}(1980)}]{bak1980}%
  \BibitemOpen
  \bibfield  {author} {\bibinfo {author} {\bibfnamefont {P.}~\bibnamefont
  {Bak}}\ and\ \bibinfo {author} {\bibfnamefont {M.~H.}\ \bibnamefont
  {Jensen}},\ }\bibfield  {title} {\enquote {\bibinfo {title} {{Theory of
  helical magnetic structures and phase transitions in MnSi and FeGe}},}\
  }\href@noop {} {\bibfield  {journal} {\bibinfo  {journal} {Journal of Physics
  C: Solid State Physics}\ }\textbf {\bibinfo {volume} {13}},\ \bibinfo {pages}
  {L881} (\bibinfo {year} {1980})}\BibitemShut {NoStop}%
\bibitem [{\citenamefont {Plumer}\ and\ \citenamefont
  {Walker}(1981)}]{plumer1981}%
  \BibitemOpen
  \bibfield  {author} {\bibinfo {author} {\bibfnamefont {M.~L.}\ \bibnamefont
  {Plumer}}\ and\ \bibinfo {author} {\bibfnamefont {M.~B.}\ \bibnamefont
  {Walker}},\ }\bibfield  {title} {\enquote {\bibinfo {title} {{Wavevector and
  spin reorientation in MnSi}},}\ }\href@noop {} {\bibfield  {journal}
  {\bibinfo  {journal} {Journal of Physics C: Solid State Physics}\ }\textbf
  {\bibinfo {volume} {14}},\ \bibinfo {pages} {4689} (\bibinfo {year}
  {1981})}\BibitemShut {NoStop}%
\bibitem [{\citenamefont {Bauer}\ \emph {et~al.}(2017)\citenamefont {Bauer},
  \citenamefont {Chacon}, \citenamefont {Wagner}, \citenamefont {Halder},
  \citenamefont {Georgii}, \citenamefont {Rosch}, \citenamefont {Pfleiderer},\
  and\ \citenamefont {Garst}}]{bauer2017}%
  \BibitemOpen
  \bibfield  {author} {\bibinfo {author} {\bibfnamefont {A.}~\bibnamefont
  {Bauer}}, \bibinfo {author} {\bibfnamefont {A.}~\bibnamefont {Chacon}},
  \bibinfo {author} {\bibfnamefont {M.}~\bibnamefont {Wagner}}, \bibinfo
  {author} {\bibfnamefont {M.}~\bibnamefont {Halder}}, \bibinfo {author}
  {\bibfnamefont {R.}~\bibnamefont {Georgii}}, \bibinfo {author} {\bibfnamefont
  {A.}~\bibnamefont {Rosch}}, \bibinfo {author} {\bibfnamefont {C.}~\bibnamefont
  {Pfleiderer}}, \ and\ \bibinfo {author} {\bibfnamefont {M.}~\bibnamefont
  {Garst}},\ }\bibfield  {title} {\enquote {\bibinfo {title} {{Symmetry
  breaking, slow relaxation dynamics, and topological defects at the
  field-induced helix reorientation in MnSi}},}\ }\href@noop {} {\bibfield
  {journal} {\bibinfo  {journal} {Phys. Rev. B}\ }\textbf {\bibinfo {volume}
  {95}},\ \bibinfo {pages} {024429} (\bibinfo {year} {2017})}\BibitemShut
  {NoStop}%
\bibitem [{\citenamefont {Maleyev}(2006)}]{maleyev2006}%
  \BibitemOpen
  \bibfield  {author} {\bibinfo {author} {\bibfnamefont {S.~V.}\ \bibnamefont
  {Maleyev}},\ }\bibfield  {title} {\enquote {\bibinfo {title} {{Cubic magnets
  with Dzyaloshinskii-Moriya interaction at low temperature}},}\ }\href@noop {}
  {\bibfield  {journal} {\bibinfo  {journal} {Phys. Rev. B}\ }\textbf {\bibinfo
  {volume} {73}},\ \bibinfo {pages} {174402} (\bibinfo {year}
  {2006})}\BibitemShut {NoStop}%
\bibitem [{\citenamefont {Milde}\ \emph {et~al.}(2020)\citenamefont {Milde},
  \citenamefont {Kohler}, \citenamefont {Neuber}, \citenamefont
  {Ritzinger}, \citenamefont {Garst}, \citenamefont {Bauer}, \citenamefont
  {Pfleiderer}, \citenamefont {Berger},\ and\ \citenamefont {Eng}}]{Milde2020}%
  \BibitemOpen
  
  \bibfield  {author} {\bibinfo {author} {\bibfnamefont {P.}~\bibnamefont
  {Milde}}, \bibinfo {author} {\bibfnamefont {L.}~\bibnamefont
  {K\"{o}hler}}, \bibinfo {author} {\bibfnamefont {E.}~\bibnamefont
  {Neuber}}, \bibinfo {author} {\bibfnamefont {P.}~\bibnamefont {Ritzinger}},
  \bibinfo {author} {\bibfnamefont {M.}~\bibnamefont {Garst}}, \bibinfo
  {author} {\bibfnamefont {A.}~\bibnamefont {Bauer}}, \bibinfo {author}
  {\bibfnamefont {C.}~\bibnamefont {Pfleiderer}}, \bibinfo {author}
  {\bibfnamefont {H.}~\bibnamefont {Berger}}, \ and\ \bibinfo {author}
  {\bibfnamefont {L.~M.}\ \bibnamefont {Eng}},\ }\bibfield  {title} {\enquote
  {\bibinfo {title} {{Field-induced reorientation of helimagnetic order in
  Cu$_2$OSeO$_3$ probed by magnetic force microscopy}},}\ }\href@noop {}
  {\bibfield  {journal} {\bibinfo  {journal} {Physical Review B}\ \textbf {\bibinfo {volume} {102}},
  \ \bibinfo {pages} {024426}} (\bibinfo {year} {2020})}\BibitemShut {NoStop}%
\bibitem [{\citenamefont {Buhrandt}\ and\ \citenamefont
  {Fritz}(2013)}]{buhrandt2013}%
  \BibitemOpen
  \bibfield  {author} {\bibinfo {author} {\bibfnamefont {S.}~\bibnamefont
  {Buhrandt}}\ and\ \bibinfo {author} {\bibfnamefont {L.}~\bibnamefont
  {Fritz}},\ }\bibfield  {title} {\enquote {\bibinfo {title} {{Skyrmion lattice
  phase in three-dimensional chiral magnets from Monte Carlo simulations}},}\
  }\href@noop {} {\bibfield  {journal} {\bibinfo  {journal} {Phys. Rev. B}\
  }\textbf {\bibinfo {volume} {88}},\ \bibinfo {pages} {195137} (\bibinfo
  {year} {2013})}\BibitemShut {NoStop}%
\bibitem [{\citenamefont {Seki}\ \emph {et~al.}(2017)\citenamefont {Seki},
  \citenamefont {Okamura}, \citenamefont {Shibata}, \citenamefont {Takagi},
  \citenamefont {Khanh}, \citenamefont {Kagawa}, \citenamefont {Arima},\ and\
  \citenamefont {Tokura}}]{seki2017uniaxial}%
  \BibitemOpen
  \bibfield  {author} {\bibinfo {author} {\bibfnamefont {S.}~\bibnamefont
  {Seki}}, \bibinfo {author} {\bibfnamefont {Y.}~\bibnamefont {Okamura}},
  \bibinfo {author} {\bibfnamefont {K.}~\bibnamefont {Shibata}}, \bibinfo
  {author} {\bibfnamefont {R.}~\bibnamefont {Takagi}}, \bibinfo {author}
  {\bibfnamefont {N.~D.}\ \bibnamefont {Khanh}}, \bibinfo {author}
  {\bibfnamefont {F.}~\bibnamefont {Kagawa}}, \bibinfo {author} {\bibfnamefont
  {T.}~\bibnamefont {Arima}}, \ and\ \bibinfo {author} {\bibfnamefont
  {Y.}~\bibnamefont {Tokura}},\ }\bibfield  {title} {\enquote {\bibinfo {title}
  {Stabilization of magnetic skyrmions by uniaxial tensile strain},}\
  }\href@noop {} {\bibfield  {journal} {\bibinfo  {journal} {Phys. Rev. B}\
  }\textbf {\bibinfo {volume} {96}},\ \bibinfo {pages} {220404(R)} (\bibinfo
  {year} {2017})}\BibitemShut {NoStop}%
\bibitem [{\citenamefont {Nakajima}\ \emph {et~al.}(2018)\citenamefont
  {Nakajima}, \citenamefont {Ukleev}, \citenamefont {Ohishi}, \citenamefont
  {Oike}, \citenamefont {Kagawa}, \citenamefont {Seki}, \citenamefont
  {Kakurai}, \citenamefont {Tokura},\ and\ \citenamefont
  {Arima}}]{nakajima2018}%
  \BibitemOpen
  \bibfield  {author} {\bibinfo {author} {\bibfnamefont {T.}~\bibnamefont
  {Nakajima}}, \bibinfo {author} {\bibfnamefont {V.}~\bibnamefont {Ukleev}},
  \bibinfo {author} {\bibfnamefont {K.}~\bibnamefont {Ohishi}}, \bibinfo
  {author} {\bibfnamefont {H.}~\bibnamefont {Oike}}, \bibinfo {author}
  {\bibfnamefont {F.}~\bibnamefont {Kagawa}}, \bibinfo {author} {\bibfnamefont
  {S.}~\bibnamefont {Seki}}, \bibinfo {author} {\bibfnamefont {K.}~\bibnamefont
  {Kakurai}}, \bibinfo {author} {\bibfnamefont {Y.}~\bibnamefont {Tokura}}, \
  and\ \bibinfo {author} {\bibfnamefont {T.}~\bibnamefont {Arima}},\ }\bibfield
   {title} {\enquote {\bibinfo {title} {{Uniaxial-stress Effects on
  Helimagnetic Orders and Skyrmion Lattice in Cu$_{2}$OSeO$_{3}$}},}\
  }\href@noop {} {\bibfield  {journal} {\bibinfo  {journal} {Journal of the
  Physical Society of Japan}\ }\textbf {\bibinfo {volume} {87}},\ \bibinfo
  {pages} {094709} (\bibinfo {year} {2018})}\BibitemShut {NoStop}%
\bibitem [{\citenamefont {Levati{\'c}}\ \emph {et~al.}(2016)\citenamefont
  {Levati{\'c}}, \citenamefont {Pop{\v c}evi{\'c}}, \citenamefont {{\v
  S}urija}, \citenamefont {Kruchkov}, \citenamefont {Berger}, \citenamefont
  {Magrez}, \citenamefont {White}, \citenamefont {R{\o}nnow},\ and\
  \citenamefont {{\v Z}ivkovi{\'c}}}]{levatic2016}%
  \BibitemOpen
  \bibfield  {author} {\bibinfo {author} {\bibfnamefont {I.}~\bibnamefont
  {Levati{\'c}}}, \bibinfo {author} {\bibfnamefont {P.}~\bibnamefont {Pop{\v
  c}evi{\'c}}}, \bibinfo {author} {\bibfnamefont {V.}~\bibnamefont {{\v
  S}urija}}, \bibinfo {author} {\bibfnamefont {A.}~\bibnamefont {Kruchkov}},
  \bibinfo {author} {\bibfnamefont {H.}~\bibnamefont {Berger}}, \bibinfo
  {author} {\bibfnamefont {A.}~\bibnamefont {Magrez}}, \bibinfo {author}
  {\bibfnamefont {J.~S.}\ \bibnamefont {White}}, \bibinfo {author}
  {\bibfnamefont {H.~M.}\ \bibnamefont {R{\o}nnow}}, \ and\ \bibinfo {author}
  {\bibfnamefont {I.}~\bibnamefont {{\v Z}ivkovi{\'c}}},\ }\bibfield  {title}
  {\enquote {\bibinfo {title} {{Dramatic pressure-driven enhancement of bulk
  skyrmion stability}},}\ }\href@noop {} {\bibfield  {journal} {\bibinfo
  {journal} {Scientific Reports}\ }\textbf {\bibinfo {volume} {6}},\ \bibinfo
  {pages} {21347} (\bibinfo {year} {2016})}\BibitemShut {NoStop}%
\bibitem [{\citenamefont {Crisanti}\ \emph {et~al.}(2020)\citenamefont
  {Crisanti}, \citenamefont {Reynolds}, \citenamefont {\ifmmode \check{Z}\else
  \v{Z}\fi{}ivkovi\ifmmode~\acute{c}\else \'{c}\fi{}}, \citenamefont {Magrez},
  \citenamefont {R\o{}nnow}, \citenamefont {Cubitt},\ and\ \citenamefont
  {White}}]{Crisanti2020}%
  \BibitemOpen
  \bibfield  {author} {\bibinfo {author} {\bibfnamefont {M.}~\bibnamefont
  {Crisanti}}, \bibinfo {author} {\bibfnamefont {N.}~\bibnamefont {Reynolds}},
  \bibinfo {author} {\bibfnamefont {I.}~\bibnamefont {\ifmmode \check{Z}\else
  \v{Z}\fi{}ivkovi\ifmmode~\acute{c}\else \'{c}\fi{}}}, \bibinfo {author}
  {\bibfnamefont {A.}~\bibnamefont {Magrez}}, \bibinfo {author} {\bibfnamefont
  {H.~M.}\ \bibnamefont {R\o{}nnow}}, \bibinfo {author} {\bibfnamefont
  {R.}~\bibnamefont {Cubitt}}, \ and\ \bibinfo {author} {\bibfnamefont {J.~S.}\
  \bibnamefont {White}},\ }\bibfield  {title} {\enquote {\bibinfo {title} {In
  situ control of the helical and skyrmion phases in cu$_2$oseo$_3$ using
  high-pressure helium gas up to 5 kbar},}\ }\href@noop {} {\bibfield
  {journal} {\bibinfo  {journal} {Phys. Rev. B}\ }\textbf {\bibinfo {volume}
  {101}},\ \bibinfo {pages} {214435} (\bibinfo {year} {2020})}\BibitemShut
  {NoStop}%
\bibitem [{\citenamefont {White}\ \emph {et~al.}(2014)\citenamefont {White},
  \citenamefont {Pr{\v{s}}a}, \citenamefont {Huang}, \citenamefont {Omrani},
  \citenamefont {{\v{Z}}ivkovi{\'c}}, \citenamefont {Bartkowiak}, \citenamefont
  {Berger}, \citenamefont {Magrez}, \citenamefont {Gavilano}, \citenamefont
  {Nagy}, \citenamefont {Zang},\ and\ \citenamefont {R{\o}nnow}}]{white2014}%
  \BibitemOpen
  \bibfield  {author} {\bibinfo {author} {\bibfnamefont {J.~S.}\ \bibnamefont
  {White}}, \bibinfo {author} {\bibfnamefont {K.}~\bibnamefont {Pr{\v{s}}a}},
  \bibinfo {author} {\bibfnamefont {P.}~\bibnamefont {Huang}}, \bibinfo
  {author} {\bibfnamefont {A.~A.}\ \bibnamefont {Omrani}}, \bibinfo {author}
  {\bibfnamefont {I.}~\bibnamefont {{\v{Z}}ivkovi{\'c}}}, \bibinfo {author}
  {\bibfnamefont {M.}~\bibnamefont {Bartkowiak}}, \bibinfo {author}
  {\bibfnamefont {H.}~\bibnamefont {Berger}}, \bibinfo {author} {\bibfnamefont
  {A.}~\bibnamefont {Magrez}}, \bibinfo {author} {\bibfnamefont {J.~L.}\
  \bibnamefont {Gavilano}}, \bibinfo {author} {\bibfnamefont {G.}~\bibnamefont
  {Nagy}}, \bibinfo {author} {\bibfnamefont {J.}~\bibnamefont {Zang}}, \ and\
  \bibinfo {author} {\bibfnamefont {H.~M.}\ \bibnamefont
  {R{\o}nnow}},\ }\bibfield  {title} {\enquote {\bibinfo {title}
  {{Electric-field-induced skyrmion distortion and giant lattice rotation in
  the magnetoelectric insulator Cu$_2$OSeO$_3$}},}\ }\href@noop {} {\bibfield
  {journal} {\bibinfo  {journal} {Phys. Rev. Lett.}\ }\textbf {\bibinfo
  {volume} {113}},\ \bibinfo {pages} {107203} (\bibinfo {year}
  {2014})}\BibitemShut {NoStop}%
\bibitem [{\citenamefont {Okamura}\ \emph {et~al.}(2016)\citenamefont
  {Okamura}, \citenamefont {Kagawa}, \citenamefont {Seki},\ and\ \citenamefont
  {Tokura}}]{okamura2016}%
  \BibitemOpen
  \bibfield  {author} {\bibinfo {author} {\bibfnamefont {Y.}~\bibnamefont
  {Okamura}}, \bibinfo {author} {\bibfnamefont {F.}~\bibnamefont {Kagawa}},
  \bibinfo {author} {\bibfnamefont {S.}~\bibnamefont {Seki}}, \ and\ \bibinfo
  {author} {\bibfnamefont {Y.}~\bibnamefont {Tokura}},\ }\bibfield  {title}
  {\enquote {\bibinfo {title} {{Transition to and from the skyrmion lattice
  phase by electric fields in a magnetoelectric compound}},}\ }\href@noop {}
  {\bibfield  {journal} {\bibinfo  {journal} {Nature Communications}\ }\textbf
  {\bibinfo {volume} {7}},\ \bibinfo {pages} {12669} (\bibinfo {year}
  {2016})}\BibitemShut {NoStop}%
\bibitem [{\citenamefont {Kruchkov}\ \emph {et~al.}(2018)\citenamefont
  {Kruchkov}, \citenamefont {White}, \citenamefont {Bartkowiak}, \citenamefont
  {{\v Z}ivkovi{\'c}}, \citenamefont {Magrez},\ and\ \citenamefont
  {R{\o}nnow}}]{kruchkov2018}%
  \BibitemOpen
  \bibfield  {author} {\bibinfo {author} {\bibfnamefont {A.~J.}\ \bibnamefont
  {Kruchkov}}, \bibinfo {author} {\bibfnamefont {J.~S.}\ \bibnamefont {White}},
  \bibinfo {author} {\bibfnamefont {M.}~\bibnamefont {Bartkowiak}}, \bibinfo
  {author} {\bibfnamefont {I.}~\bibnamefont {{\v Z}ivkovi{\'c}}}, \bibinfo
  {author} {\bibfnamefont {A.}~\bibnamefont {Magrez}}, \ and\ \bibinfo {author}
  {\bibfnamefont {H.~M.}\ \bibnamefont {R{\o}nnow}},\ }\bibfield  {title}
  {\enquote {\bibinfo {title} {{Direct electric field control of the skyrmion
  phase in a magnetoelectric insulator}},}\ }\href@noop {} {\bibfield
  {journal} {\bibinfo  {journal} {Scientific Reports}\ }\textbf {\bibinfo
  {volume} {8}},\ \bibinfo {pages} {10466} (\bibinfo {year}
  {2018})}\BibitemShut {NoStop}%
\bibitem [{\citenamefont {White}\ \emph {et~al.}(2018)\citenamefont {White},
  \citenamefont {{\v{Z}}ivkovi{\'c}}, \citenamefont {Kruchkov}, \citenamefont
  {Bartkowiak}, \citenamefont {Magrez},\ and\ \citenamefont
  {R{\o}nnow}}]{white2018}%
  \BibitemOpen
  \bibfield  {author} {\bibinfo {author} {\bibfnamefont {J.~S.}\ \bibnamefont
  {White}}, \bibinfo {author} {\bibfnamefont {I.}~\bibnamefont
  {{\v{Z}}ivkovi{\'c}}}, \bibinfo {author} {\bibfnamefont {A.~J.}\ \bibnamefont
  {Kruchkov}}, \bibinfo {author} {\bibfnamefont {M.}~\bibnamefont
  {Bartkowiak}}, \bibinfo {author} {\bibfnamefont {A.}~\bibnamefont {Magrez}},
  \ and\ \bibinfo {author} {\bibfnamefont {H.~M.}\ \bibnamefont {R{\o}nnow}},\
  }\bibfield  {title} {\enquote {\bibinfo {title} {{Electric-Field-Driven
  Topological Phase Switching and Skyrmion-Lattice Metastability in
  Magnetoelectric Cu$_{2}$OSeO$_{3}$}},}\ }\href@noop {} {\bibfield  {journal}
  {\bibinfo  {journal} {Phys. Rev. Applied}\ }\textbf {\bibinfo {volume}
  {10}},\ \bibinfo {pages} {014021} (\bibinfo {year} {2018})}\BibitemShut
  {NoStop}%
\bibitem [{\citenamefont {{\v{S}}tefan{\v{c}}i{\v{c}}}\ \emph
  {et~al.}(2018)\citenamefont {{\v{S}}tefan{\v{c}}i{\v{c}}}, \citenamefont
  {Moody}, \citenamefont {Hicken}, \citenamefont {Birch}, \citenamefont
  {Balakrishnan}, \citenamefont {Barnett}, \citenamefont {Crisanti},
  \citenamefont {Evans}, \citenamefont {Holt}, \citenamefont {Franke},
  \citenamefont {Hatton}, \citenamefont {Huddart}, \citenamefont {Lees},
  \citenamefont {Pratt}, \citenamefont {Tang}, \citenamefont {Wilson},
  \citenamefont {Xiao},\ and\ \citenamefont {Lancaster}}]{stefanvcivc2018}%
  \BibitemOpen
  \bibfield  {author} {\bibinfo {author} {\bibfnamefont {A.}~\bibnamefont
  {{\v{S}}tefan{\v{c}}i{\v{c}}}}, \bibinfo {author} {\bibfnamefont {S.~H.}\
  \bibnamefont {Moody}}, \bibinfo {author} {\bibfnamefont {T.~J.}\ \bibnamefont
  {Hicken}}, \bibinfo {author} {\bibfnamefont {M.~T.}\ \bibnamefont {Birch}},
  \bibinfo {author} {\bibfnamefont {G.}~\bibnamefont {Balakrishnan}}, \bibinfo
  {author} {\bibfnamefont {S.~A.}\ \bibnamefont {Barnett}}, \bibinfo {author}
  {\bibfnamefont {M.}~\bibnamefont {Crisanti}}, \bibinfo {author}
  {\bibfnamefont {J.~S.~O.}\ \bibnamefont {Evans}}, \bibinfo {author}
  {\bibfnamefont {S.~J.~R.}\ \bibnamefont {Holt}}, \bibinfo {author}
  {\bibfnamefont {K.~J.~A.}\ \bibnamefont {Franke}}, \bibinfo {author}
  {\bibfnamefont {P.~D.}\ \bibnamefont {Hatton}}, \bibinfo {author}
  {\bibfnamefont {B.~M.}\ \bibnamefont {Huddart}}, \bibinfo {author}
  {\bibfnamefont {M.~R.}\ \bibnamefont {Lees}}, \bibinfo {author}
  {\bibfnamefont {F.~L.}\ \bibnamefont {Pratt}}, \bibinfo {author}
  {\bibfnamefont {C.~C.}\ \bibnamefont {Tang}}, \bibinfo {author}
  {\bibfnamefont {M.~N.}\ \bibnamefont {Wilson}}, \bibinfo {author}
  {\bibfnamefont {F.}~\bibnamefont {Xiao}}, \ and\ \bibinfo {author}
  {\bibfnamefont {T.}~\bibnamefont {Lancaster}},\ }\bibfield  {title} {\enquote
  {\bibinfo {title} {{Origin of skyrmion lattice phase splitting in
  Zn-substituted Cu$_2$OSeO$_3$}},}\ }\href@noop {} {\bibfield  {journal}
  {\bibinfo  {journal} {Phys. Rev. Materials}\ }\textbf {\bibinfo {volume}
  {2}},\ \bibinfo {pages} {111402(R)} (\bibinfo {year} {2018})}\BibitemShut
  {NoStop}%
\bibitem [{\citenamefont {Sukhanov}\ \emph {et~al.}(2019)\citenamefont
  {Sukhanov}, \citenamefont {Vir}, \citenamefont {Cameron}, \citenamefont {Wu},
  \citenamefont {Martin}, \citenamefont {M\"uhlbauer}, \citenamefont
  {Heinemann}, \citenamefont {Yang}, \citenamefont {Felser},\ and\
  \citenamefont {Inosov}}]{Sukhanov2019}%
  \BibitemOpen
  \bibfield  {author} {\bibinfo {author} {\bibfnamefont {A.~S.}\ \bibnamefont
  {Sukhanov}}, \bibinfo {author} {\bibfnamefont {Praveen}\ \bibnamefont {Vir}},
  \bibinfo {author} {\bibfnamefont {A.~S.}\ \bibnamefont {Cameron}}, \bibinfo
  {author} {\bibfnamefont {H.~C.}\ \bibnamefont {Wu}}, \bibinfo {author}
  {\bibfnamefont {N.}~\bibnamefont {Martin}}, \bibinfo {author} {\bibfnamefont
  {S.}~\bibnamefont {M\"uhlbauer}}, \bibinfo {author} {\bibfnamefont
  {A.}~\bibnamefont {Heinemann}}, \bibinfo {author} {\bibfnamefont {H.~D.}\
  \bibnamefont {Yang}}, \bibinfo {author} {\bibfnamefont {C.}~\bibnamefont
  {Felser}}, \ and\ \bibinfo {author} {\bibfnamefont {D.~S.}\ \bibnamefont
  {Inosov}},\ }\bibfield  {title} {\enquote {\bibinfo {title} {Increasing
  skyrmion stability in ${\mathrm{cu}}_{2}{\mathrm{oseo}}_{3}$ by chemical
  substitution},}\ }\href@noop {} {\bibfield  {journal} {\bibinfo  {journal}
  {Phys. Rev. B}\ }\textbf {\bibinfo {volume} {100}},\ \bibinfo {pages}
  {184408} (\bibinfo {year} {2019})}\BibitemShut {NoStop}%
\bibitem [{\citenamefont {Bogdanov}\ and\ \citenamefont
  {Yablonskii}(1989)}]{bogdanov1989}%
  \BibitemOpen
  \bibfield  {author} {\bibinfo {author} {\bibfnamefont {A.~N.}\ \bibnamefont
  {Bogdanov}}\ and\ \bibinfo {author} {\bibfnamefont {D.~A.}\ \bibnamefont
  {Yablonskii}},\ }\bibfield  {title} {\enquote {\bibinfo {title}
  {Thermodynamically stable `vortices' in magnetically ordered crystals. the
  mixed state of magnets},}\ }\href@noop {} {\bibfield  {journal} {\bibinfo
  {journal} {Zh. Eksp. Teor. Fiz}\ }\textbf {\bibinfo {volume} {95}},\ \bibinfo
  {pages} {182} (\bibinfo {year} {1989})}\BibitemShut {NoStop}%
\bibitem [{\citenamefont {Bogdanov}\ and\ \citenamefont
  {Hubert}(1999)}]{bogdanov1999}%
  \BibitemOpen
  \bibfield  {author} {\bibinfo {author} {\bibfnamefont {A.}~\bibnamefont
  {Bogdanov}}\ and\ \bibinfo {author} {\bibfnamefont {A.}~\bibnamefont
  {Hubert}},\ }\bibfield  {title} {\enquote {\bibinfo {title} {{The Stability
  of Vortex-Like Structures in Uniaxial Ferromagnets}},}\ }\href@noop {}
  {\bibfield  {journal} {\bibinfo  {journal} {Journal of Magnetism and Magnetic
  Materials}\ }\textbf {\bibinfo {volume} {195}},\ \bibinfo {pages} {182}
  (\bibinfo {year} {1999})}\BibitemShut {NoStop}%
\bibitem [{\citenamefont {R{\"o}{\ss}ler}\ \emph {et~al.}(2006)\citenamefont
  {R{\"o}{\ss}ler}, \citenamefont {Bogdanov},\ and\ \citenamefont
  {Pfleiderer}}]{rossler2006}%
  \BibitemOpen
  \bibfield  {author} {\bibinfo {author} {\bibfnamefont {U.~K.}\ \bibnamefont
  {R{\"o}{\ss}ler}}, \bibinfo {author} {\bibfnamefont {A.~N.}\ \bibnamefont
  {Bogdanov}}, \ and\ \bibinfo {author} {\bibfnamefont {C.}~\bibnamefont
  {Pfleiderer}},\ }\bibfield  {title} {\enquote {\bibinfo {title} {{Spontaneous
  skyrmion ground states in magnetic metals}},}\ }\href@noop {} {\bibfield
  {journal} {\bibinfo  {journal} {Nature}\ }\textbf {\bibinfo {volume} {442}},\
  \bibinfo {pages} {797--801} (\bibinfo {year} {2006})}\BibitemShut {NoStop}%
\bibitem [{\citenamefont {Leonov}(2012)}]{leonovPHD}%
  \BibitemOpen
  \bibfield  {author} {\bibinfo {author} {\bibfnamefont {A.~O.}\ \bibnamefont
  {Leonov}},\ }\emph {\bibinfo {title} {Twisted, localized, and modulated
  states described in the phenomenological theory of chiral and nanoscale
  ferromagnets}},\ \href@noop {} {Ph.D. thesis},\ \bibinfo  {school} {Technical
  University of Dresden} (\bibinfo {year} {2012})\BibitemShut {NoStop}%
\bibitem [{\citenamefont {Butenko}\ \emph {et~al.}(2010)\citenamefont
  {Butenko}, \citenamefont {Leonov}, \citenamefont {R{\"o}{\ss}ler},\ and\
  \citenamefont {Bogdanov}}]{butenko2010}%
  \BibitemOpen
  \bibfield  {author} {\bibinfo {author} {\bibfnamefont {A.~B.}\ \bibnamefont
  {Butenko}}, \bibinfo {author} {\bibfnamefont {A.~A.}\ \bibnamefont {Leonov}},
  \bibinfo {author} {\bibfnamefont {U.~K.}\ \bibnamefont {R{\"o}{\ss}ler}}, \
  and\ \bibinfo {author} {\bibfnamefont {A.~N.}\ \bibnamefont {Bogdanov}},\
  }\bibfield  {title} {\enquote {\bibinfo {title} {{Stabilization of skyrmion
  textures by uniaxial distortions in noncentrosymmetric cubic helimagnets}},}\
  }\href@noop {} {\bibfield  {journal} {\bibinfo  {journal} {Phys. Rev. B}\
  }\textbf {\bibinfo {volume} {82}},\ \bibinfo {pages} {052403} (\bibinfo
  {year} {2010})}\BibitemShut {NoStop}%
\bibitem [{\citenamefont {Qian}\ \emph {et~al.}(2018)\citenamefont {Qian},
  \citenamefont {Bannenberg}, \citenamefont {Wilhelm}, \citenamefont
  {Chaboussant}, \citenamefont {DeBeer-Schmitt}, \citenamefont {Schmidt},
  \citenamefont {Aqeel}, \citenamefont {Palstra}, \citenamefont {Br{\"u}ck},
  \citenamefont {Lefering}, \citenamefont {Pappas}, \citenamefont {Mostovoy},\
  and\ \citenamefont {Leonov}}]{qian2018}%
  \BibitemOpen
  \bibfield  {author} {\bibinfo {author} {\bibfnamefont {F.}~\bibnamefont
  {Qian}}, \bibinfo {author} {\bibfnamefont {L.~J.}\ \bibnamefont
  {Bannenberg}}, \bibinfo {author} {\bibfnamefont {H.}~\bibnamefont {Wilhelm}},
  \bibinfo {author} {\bibfnamefont {G.}~\bibnamefont {Chaboussant}}, \bibinfo
  {author} {\bibfnamefont {L.~M.}\ \bibnamefont {DeBeer-Schmitt}}, \bibinfo
  {author} {\bibfnamefont {M.~P.}\ \bibnamefont {Schmidt}}, \bibinfo {author}
  {\bibfnamefont {A.}~\bibnamefont {Aqeel}}, \bibinfo {author} {\bibfnamefont
  {T.~T.~M.}\ \bibnamefont {Palstra}}, \bibinfo {author} {\bibfnamefont
  {E.~H.}\ \bibnamefont {Br{\"u}ck}}, \bibinfo {author} {\bibfnamefont
  {A.~J.~E.}\ \bibnamefont {Lefering}}, \bibinfo {author} {\bibfnamefont
  {C.}~\bibnamefont {Pappas}}, \bibinfo {author} {\bibfnamefont
  {M.}~\bibnamefont {Mostovoy}}, \ and\ \bibinfo {author} {\bibfnamefont
  {A.~O.}\ \bibnamefont {Leonov}},\ }\bibfield  {title} {\enquote {\bibinfo
  {title} {{New magnetic phase of the chiral skyrmion material
  Cu$_2$OSeO$_3$}},}\ }\href@noop {} {\bibfield  {journal} {\bibinfo  {journal}
  {Science Advances}\ }\textbf {\bibinfo {volume} {4}},\ \bibinfo {pages}
  {eaat7323} (\bibinfo {year} {2018})}\BibitemShut {NoStop}%
\bibitem [{\citenamefont {Chacon}\ \emph {et~al.}(2018)\citenamefont {Chacon},
  \citenamefont {Heinen}, \citenamefont {Halder}, \citenamefont {Bauer},
  \citenamefont {Simeth}, \citenamefont {M\"{u}hlbauer}, \citenamefont
  {Berger}, \citenamefont {Garst}, \citenamefont {Rosch},\ and\ \citenamefont
  {Pfleiderer}}]{chacon2018}%
  \BibitemOpen
  \bibfield  {author} {\bibinfo {author} {\bibfnamefont {A.}~\bibnamefont
  {Chacon}}, \bibinfo {author} {\bibfnamefont {L.}~\bibnamefont {Heinen}},
  \bibinfo {author} {\bibfnamefont {M.}~\bibnamefont {Halder}}, \bibinfo
  {author} {\bibfnamefont {A.}~\bibnamefont {Bauer}}, \bibinfo {author}
  {\bibfnamefont {W.}~\bibnamefont {Simeth}}, \bibinfo {author} {\bibfnamefont
  {S.}~\bibnamefont {M\"{u}hlbauer}}, \bibinfo {author} {\bibfnamefont
  {H.}~\bibnamefont {Berger}}, \bibinfo {author} {\bibfnamefont
  {M.}~\bibnamefont {Garst}}, \bibinfo {author} {\bibfnamefont
  {A.}~\bibnamefont {Rosch}}, \ and\ \bibinfo {author} {\bibfnamefont
  {C.}~\bibnamefont {Pfleiderer}},\ }\bibfield  {title} {\enquote {\bibinfo
  {title} {{Observation of two independent skyrmion phases in a chiral magnetic
  material}},}\ }\href@noop {} {\bibfield  {journal} {\bibinfo  {journal}
  {Nature Physics}\ ,\ \bibinfo {pages} {1--7}} (\bibinfo {year}
  {2018})}\BibitemShut {NoStop}%
\bibitem [{\citenamefont {Halder}\ \emph {et~al.}(2018)\citenamefont {Halder},
  \citenamefont {Chacon}, \citenamefont {Bauer}, \citenamefont {Simeth},
  \citenamefont {M\"uhlbauer}, \citenamefont {Berger}, \citenamefont {Heinen},
  \citenamefont {Garst}, \citenamefont {Rosch},\ and\ \citenamefont
  {Pfleiderer}}]{Halder2018}%
  \BibitemOpen
  \bibfield  {author} {\bibinfo {author} {\bibfnamefont {M.}~\bibnamefont
  {Halder}}, \bibinfo {author} {\bibfnamefont {A.}~\bibnamefont {Chacon}},
  \bibinfo {author} {\bibfnamefont {A.}~\bibnamefont {Bauer}}, \bibinfo
  {author} {\bibfnamefont {W.}~\bibnamefont {Simeth}}, \bibinfo {author}
  {\bibfnamefont {S.}~\bibnamefont {M\"uhlbauer}}, \bibinfo {author}
  {\bibfnamefont {H.}~\bibnamefont {Berger}}, \bibinfo {author} {\bibfnamefont
  {L.}~\bibnamefont {Heinen}}, \bibinfo {author} {\bibfnamefont
  {M.}~\bibnamefont {Garst}}, \bibinfo {author} {\bibfnamefont
  {A.}~\bibnamefont {Rosch}}, \ and\ \bibinfo {author} {\bibfnamefont
  {C.}~\bibnamefont {Pfleiderer}},\ }\bibfield  {title} {\enquote {\bibinfo
  {title} {{Thermodynamic evidence of a second skyrmion lattice phase and
  tilted conical phase in ${\mathrm{Cu}}_{2}{\mathrm{OSeO}}_{3}$}},}\
  }\href@noop {} {\bibfield  {journal} {\bibinfo  {journal} {Phys. Rev. B}\
  }\textbf {\bibinfo {volume} {98}},\ \bibinfo {pages} {144429} (\bibinfo
  {year} {2018})}\BibitemShut {NoStop}%
\bibitem [{\citenamefont {Bannenberg}\ \emph {et~al.}(2019)\citenamefont
  {Bannenberg}, \citenamefont {Wilhelm}, \citenamefont {Cubitt}, \citenamefont
  {Labh}, \citenamefont {Schmidt}, \citenamefont {Leli{\`e}vre-Berna},
  \citenamefont {Pappas}, \citenamefont {Mostovoy},\ and\ \citenamefont
  {Leonov}}]{bannenberg_npj}%
  \BibitemOpen
  \bibfield  {author} {\bibinfo {author} {\bibfnamefont {L.~J}\ \bibnamefont
  {Bannenberg}}, \bibinfo {author} {\bibfnamefont {H.}\ \bibnamefont
  {Wilhelm}}, \bibinfo {author} {\bibfnamefont {R.}\ \bibnamefont
  {Cubitt}}, \bibinfo {author} {\bibfnamefont {A.}\ \bibnamefont {Labh}},
  \bibinfo {author} {\bibfnamefont {M.~P}\ \bibnamefont {Schmidt}},
  \bibinfo {author} {\bibfnamefont {E.}\ \bibnamefont {Leli{\`e}vre-Berna}},
  \bibinfo {author} {\bibfnamefont {C.}\ \bibnamefont {Pappas}},
  \bibinfo {author} {\bibfnamefont {M.}\ \bibnamefont {Mostovoy}}, \ and\
  \bibinfo {author} {\bibfnamefont {A.~O.}\ \bibnamefont {Leonov}},\
  }\bibfield  {title} {\enquote {\bibinfo {title} {Multiple low-temperature
  skyrmionic states in a bulk chiral magnet},}\ }\href@noop {} {\bibfield
  {journal} {\bibinfo  {journal} {npj Quantum Materials}\ }\textbf {\bibinfo
  {volume} {4}},\ \bibinfo {pages} {11} (\bibinfo {year} {2019})}\BibitemShut
  {NoStop}%
\bibitem [{\citenamefont {Janson}\ \emph {et~al.}(2014)\citenamefont {Janson},
  \citenamefont {Rousochatzakis}, \citenamefont {Tsirlin}, \citenamefont
  {Belesi}, \citenamefont {Leonov}, \citenamefont {R{\"o}{\ss}ler},
  \citenamefont {van~den Brink},\ and\ \citenamefont {Rosner}}]{janson2014}%
  \BibitemOpen
  \bibfield  {author} {\bibinfo {author} {\bibfnamefont {O.}~\bibnamefont
  {Janson}}, \bibinfo {author} {\bibfnamefont {I.}~\bibnamefont
  {Rousochatzakis}}, \bibinfo {author} {\bibfnamefont {A.~A.}\ \bibnamefont
  {Tsirlin}}, \bibinfo {author} {\bibfnamefont {M.}~\bibnamefont {Belesi}},
  \bibinfo {author} {\bibfnamefont {A.~O}\ \bibnamefont {Leonov}}, \bibinfo
  {author} {\bibfnamefont {U.~K.}\ \bibnamefont {R{\"o}{\ss}ler}}, \bibinfo
  {author} {\bibfnamefont {J.}~\bibnamefont {van~den Brink}}, \ and\ \bibinfo
  {author} {\bibfnamefont {H.}~\bibnamefont {Rosner}},\ }\bibfield  {title}
  {\enquote {\bibinfo {title} {{The Quantum Nature of Skyrmions and
  Half-Skyrmions in Cu$_{2}$OSeO$_{3}$}},}\ }\href@noop {} {\bibfield
  {journal} {\bibinfo  {journal} {Nature Commununications}\ }\textbf {\bibinfo
  {volume} {5}} (\bibinfo {year} {2014})}\BibitemShut {NoStop}%
\bibitem [{\citenamefont {Dzyaloshinskii}(1965)}]{Dzyaloshinskii1965a}%
  \BibitemOpen
  \bibfield  {author} {\bibinfo {author} {\bibfnamefont {I.~E.}\ \bibnamefont
  {Dzyaloshinskii}},\ }\bibfield  {title} {\enquote {\bibinfo {title} {{The
  theory of Helicoidal Structures in Antiferromagnets. II. Metals}},}\
  }\href@noop {} {\bibfield  {journal} {\bibinfo  {journal} {Soviet Physics
  JETP}\ }\textbf {\bibinfo {volume} {20}},\ \bibinfo {pages} {665} (\bibinfo
  {year} {1965})}\BibitemShut {NoStop}%
\bibitem [{\citenamefont {Leonov}\ and\ \citenamefont
  {Pappas}(2022)}]{leonov2022prr}%
  \BibitemOpen
  \bibfield  {author} {\bibinfo {author} {\bibfnamefont {A.O.}\ \bibnamefont
  {Leonov}}\ and\ \bibinfo {author} {\bibfnamefont {C.}~\bibnamefont
  {Pappas}},\ }\bibfield  {title} {\enquote {\bibinfo {title} {{Topological
  boundaries between helical domains as a nucleation source of skyrmions in the
  bulk cubic helimagnet Cu$_2$OSeO$_3$}},}\ }\href@noop {} {\bibfield
  {journal} {\bibinfo  {journal} {Physical Review Research}\ \textbf {\bibinfo {volume} {4}},\ \bibinfo
  {pages} {043137}} (\bibinfo {year} {2022})}\BibitemShut {NoStop}%
\bibitem [{\citenamefont {Moody}\ \emph {et~al.}(2021)\citenamefont {Moody},
  \citenamefont {Nielsen}, \citenamefont {Wilson}, \citenamefont {Venero},
  \citenamefont {\v{S}tefan\v{c}i\v{c}}, \citenamefont {Balakrishnan},\ and\
  \citenamefont {Hatton}}]{Moody2021}%
  \BibitemOpen
  \bibfield  {author} {\bibinfo {author} {\bibfnamefont {S.~H.}\ \bibnamefont
  {Moody}}, \bibinfo {author} {\bibfnamefont {P.}~\bibnamefont {Nielsen}},
  \bibinfo {author} {\bibfnamefont {M.~N.}\ \bibnamefont {Wilson}}, \bibinfo
  {author} {\bibfnamefont {D.~A.}\ \bibnamefont {Venero}}, \bibinfo {author}
  {\bibfnamefont {A.}~\bibnamefont {\v{S}tefan\v{c}i\v{c}}}, \bibinfo {author}
  {\bibfnamefont {G.}~\bibnamefont {Balakrishnan}}, \ and\ \bibinfo {author}
  {\bibfnamefont {P.~D.}\ \bibnamefont {Hatton}},\ }\bibfield  {title}
  {\enquote {\bibinfo {title} {{Experimental evidence of a change of exchange
  anisotropy sign with temperature in Zn-substituted Cu$_2$OSeO$_3$}},}\
  }\href@noop {} {\bibfield  {journal} {\bibinfo  {journal} {Physical Review
  Research}\ \textbf{\bibinfo {volume} {3}},\ \bibinfo {pages} {043149}} (\bibinfo {year}
  {2021})}\BibitemShut {NoStop}%
\bibitem [{\citenamefont {{\v{Z}}ivkovi{\'c}}\ \emph
  {et~al.}(2012)\citenamefont {{\v{Z}}ivkovi{\'c}}, \citenamefont {Paji{\'c}},
  \citenamefont {Ivek},\ and\ \citenamefont {Berger}}]{zivkovic2012}%
  \BibitemOpen
  \bibfield  {author} {\bibinfo {author} {\bibfnamefont {I.}\ \bibnamefont
  {{\v{Z}}ivkovi{\'c}}}, \bibinfo {author} {\bibfnamefont {D.}\ \bibnamefont
  {Paji{\'c}}}, \bibinfo {author} {\bibfnamefont {T.}\ \bibnamefont
  {Ivek}}, \ and\ \bibinfo {author} {\bibfnamefont {H.}\ \bibnamefont
  {Berger}},\ }\bibfield  {title} {\enquote {\bibinfo {title} {{Two-step
  transition in a magnetoelectric ferrimagnet Cu$_2$OSeO$_3$}},}\ }\href@noop
  {} {\bibfield  {journal} {\bibinfo  {journal} {Phys. Rev. B}\ }\textbf
  {\bibinfo {volume} {85}},\ \bibinfo {pages} {224402} (\bibinfo {year}
  {2012})}\BibitemShut {NoStop}%
\bibitem [{\citenamefont {Leonov}\ and\ \citenamefont
  {Smalyukh}(2021)}]{leonov2021}%
  \BibitemOpen
  \bibfield  {author} {\bibinfo {author} {\bibfnamefont {A.~O.}~\bibnamefont
  {Leonov}}, \bibinfo {author} {\bibfnamefont {C.}~\bibnamefont{Pappas}}\ and\ \bibinfo {author}
  {\bibfnamefont {I.~I.}~\bibnamefont {Smalyukh}},\ }\bibfield  {title} {\enquote
  {\bibinfo {title} {{Field-driven metamorphoses of isolated skyrmions within
  the conical state of cubic helimagnets}},}\ }\href@noop {} {\bibfield
  {journal} {\bibinfo  {journal} {Phys. Rev. B}\ }\textbf {\bibinfo {volume}
  {104}},\ \bibinfo {pages} {064432} (\bibinfo {year} {2021})}\BibitemShut
  {NoStop}%
\end{thebibliography}
%
\pagebreak
\clearpage

\appendix
\begin{appendices}

\begin{minipage}{2\linewidth}
\begin{center}
{\LARGE APPENDIX}
\end{center}
\section{Results for other temperatures than those discussed in the main text}
\begin{figure}[H]
    \centering
    \includegraphics[width=\linewidth]{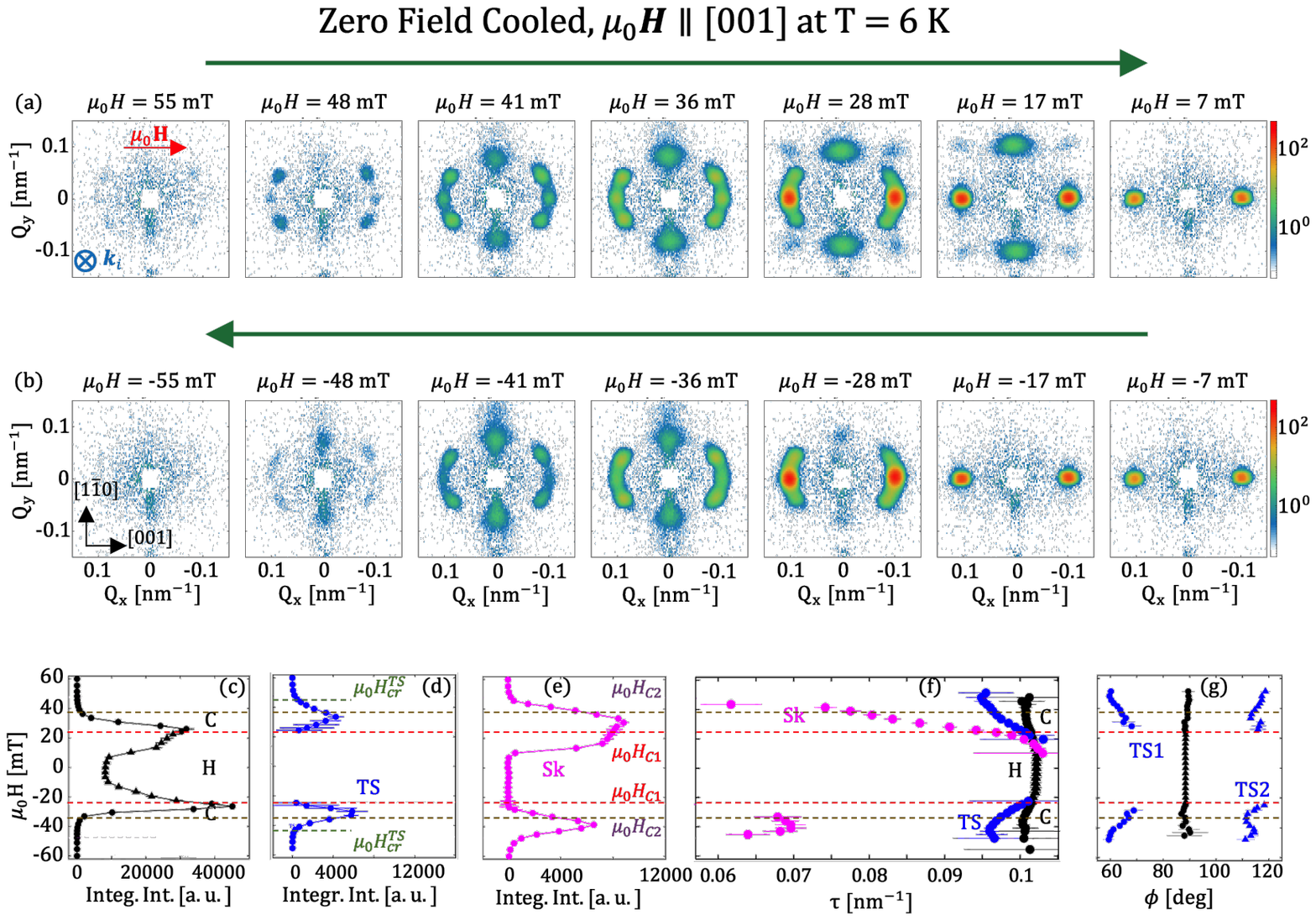}
    \caption{ (a) and (b) show SANS patterns recorded at $T = 6$~K for selected fields using the protocol schematically depicted in Figure~1 of the main text. The arrows above the SANS patterns indicate the direction of change of the magnetic field. The red arrow in the $\mu_0 H = 55$~mT panel indicates the direction of the applied magnetic field, while $k_i$ the direction of the incoming neutron beam. The field dependency of the integrated intensities of the conical (C), helical (H), TS and skyrmion peaks are shown in panels (c), (d) and (e), respectively. The corresponding values of the  characteristic wavenumbers $\tau$ are provided in (f). (g) shows the azimuthal position of the conical/helical and TS peaks. In  \textbf{(c-g)} the horizontal dashed lines  indicate the critical fields $\mu_0 H_{C1}$ and $\mu_0 H_{C2}$, respectively. The green dashed lines in (d) stand for $\mu_0 H_{cr}^{TS}$.   }
    \label{fig:panel6K}
\end{figure}
\end{minipage}

\vfill\null
\hfill
\pagebreak
\clearpage

\begin{minipage}{2\linewidth}
\begin{figure}[H]
    \centering
    \includegraphics[width=\linewidth]{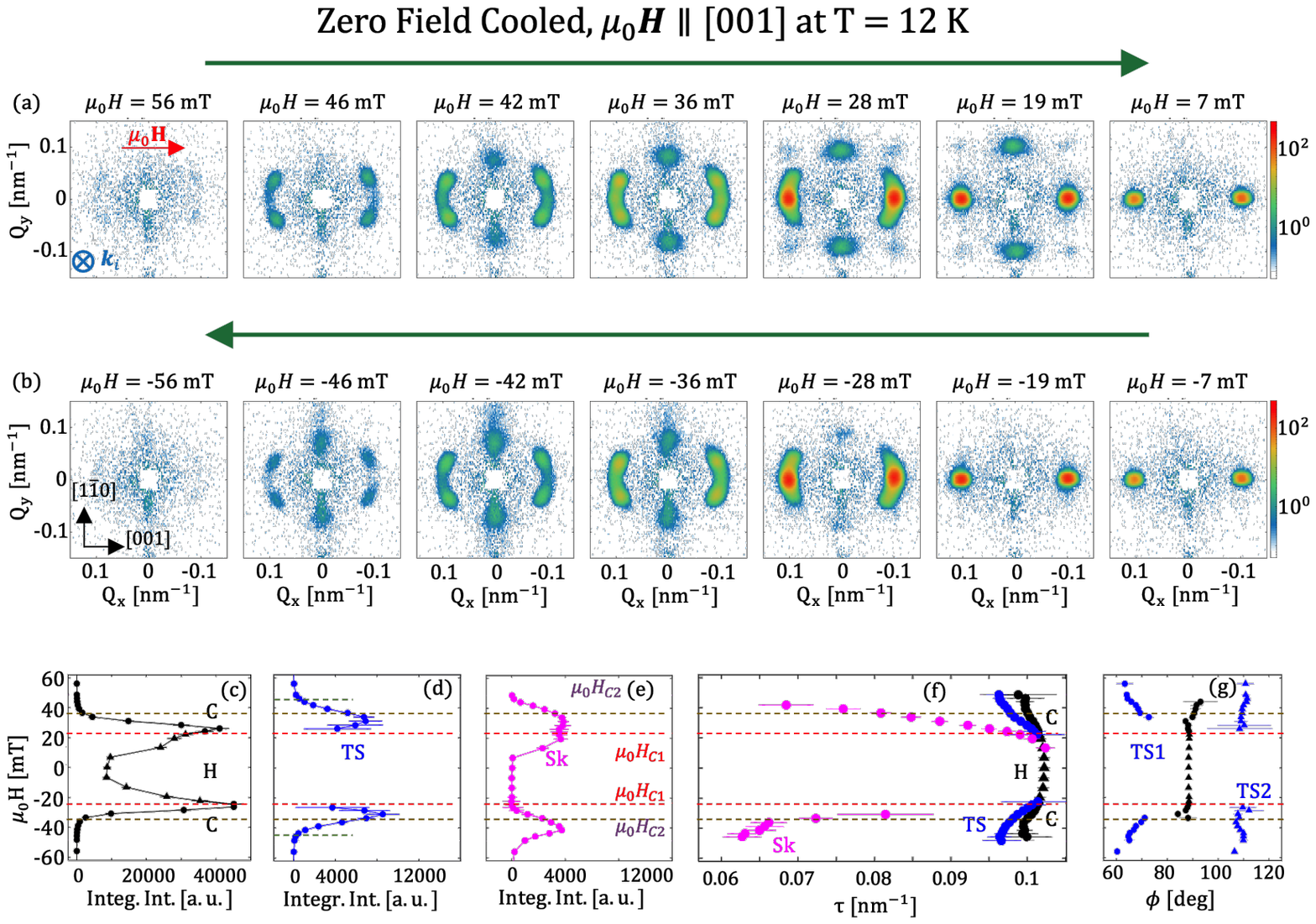}
    \caption{ (a) and (b) show SANS patterns recorded at $T = 12$~K for selected fields using the protocol schematically depicted in Figure~1 of the main text. The arrows above the SANS patterns indicate the direction of change of the magnetic field. The red arrow in the $\mu_0 H = 56$~mT panel indicates the direction of the applied magnetic field, while $k_i$ the direction of the incoming neutron beam. The field dependency of the integrated intensities of the conical (C), helical (H), TS and skyrmion peaks are shown in panels (c), (d) and (e), respectively. The corresponding values of the  characteristic wavenumbers $\tau$ are provided in (f). (g) shows the azimuthal position of the conical/helical and TS peaks. In  \textbf{(c-g)} the horizontal dashed lines  indicate the critical fields $\mu_0 H_{C1}$ and $\mu_0 H_{C2}$, respectively. The green dashed lines in (d) stand for $\mu_0 H_{cr}^{TS}$.  }
    \label{fig:panel12K}
\end{figure}
\end{minipage}
\vfill\null
\hfill
\pagebreak
\clearpage

\begin{minipage}{2\linewidth}
\begin{figure}[H]
    \centering
    \includegraphics[width=\linewidth]{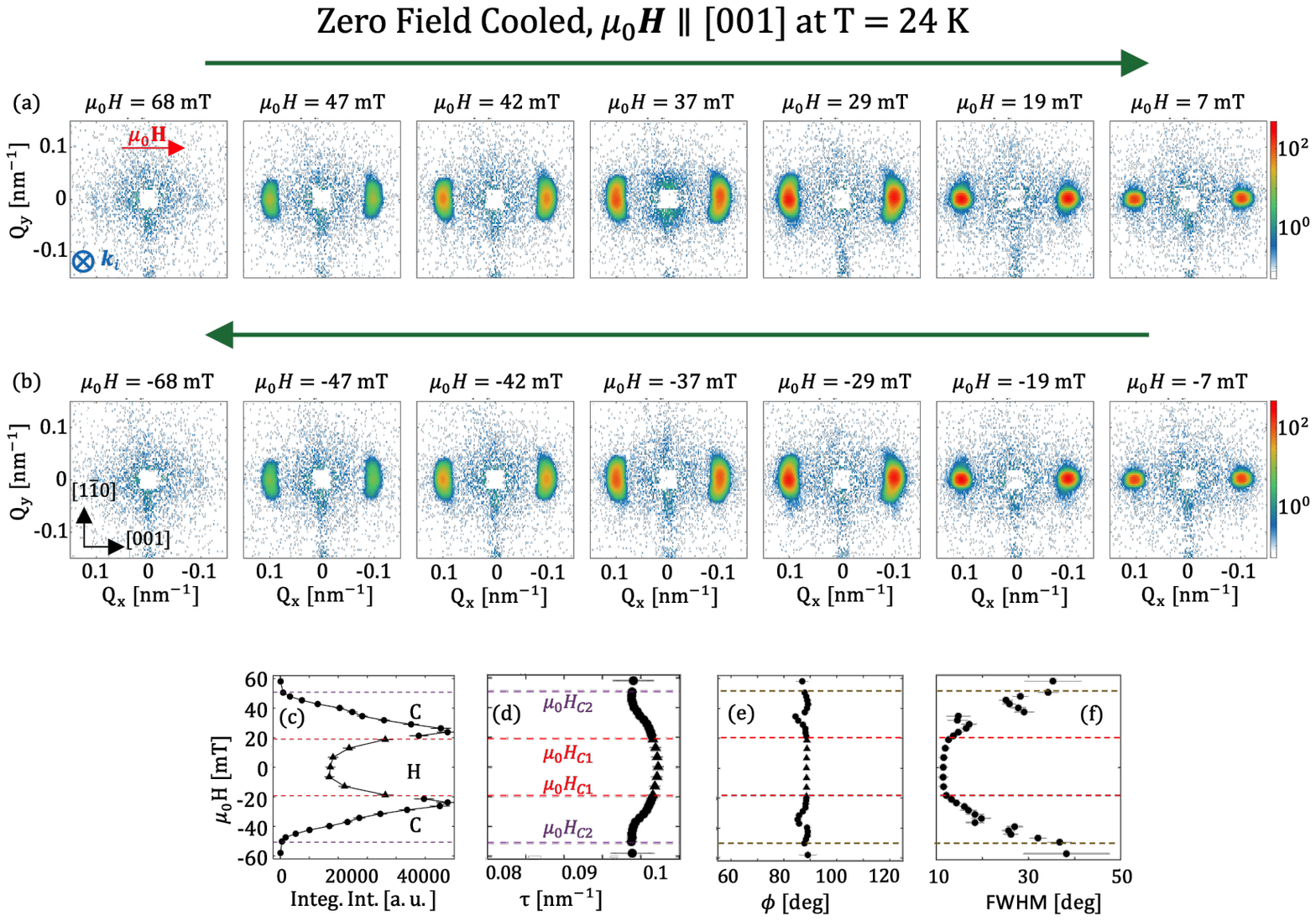}
    \caption{ (a) and (b) show SANS patterns recorded at $T = 24$~K for selected fields using the protocol schematically depicted in Figure~1 of the main text. The arrows above the SANS patterns indicate the direction of change of the magnetic field. The red arrow in the $\mu_0 H = 68$~mT panel indicates the direction of the applied magnetic field, while $k_i$ the direction of the incoming neutron beam. The field dependency of the integrated intensity is shown in (c). The values of the  characteristic wavenumbers $\tau$ are provided in (d). (e) shows the azimuthal position of the conical/helical and TS peaks and (f) their FHWM.   }
    \label{fig:panel24K}
\end{figure}
\end{minipage}
\vfill\null
\hfill

\pagebreak
\clearpage
\begin{minipage}{2\linewidth}
\section{Contour Plot for T=35 K}

\begin{figure}[H]
    \centering
    \includegraphics[width = 0.6 \textwidth]{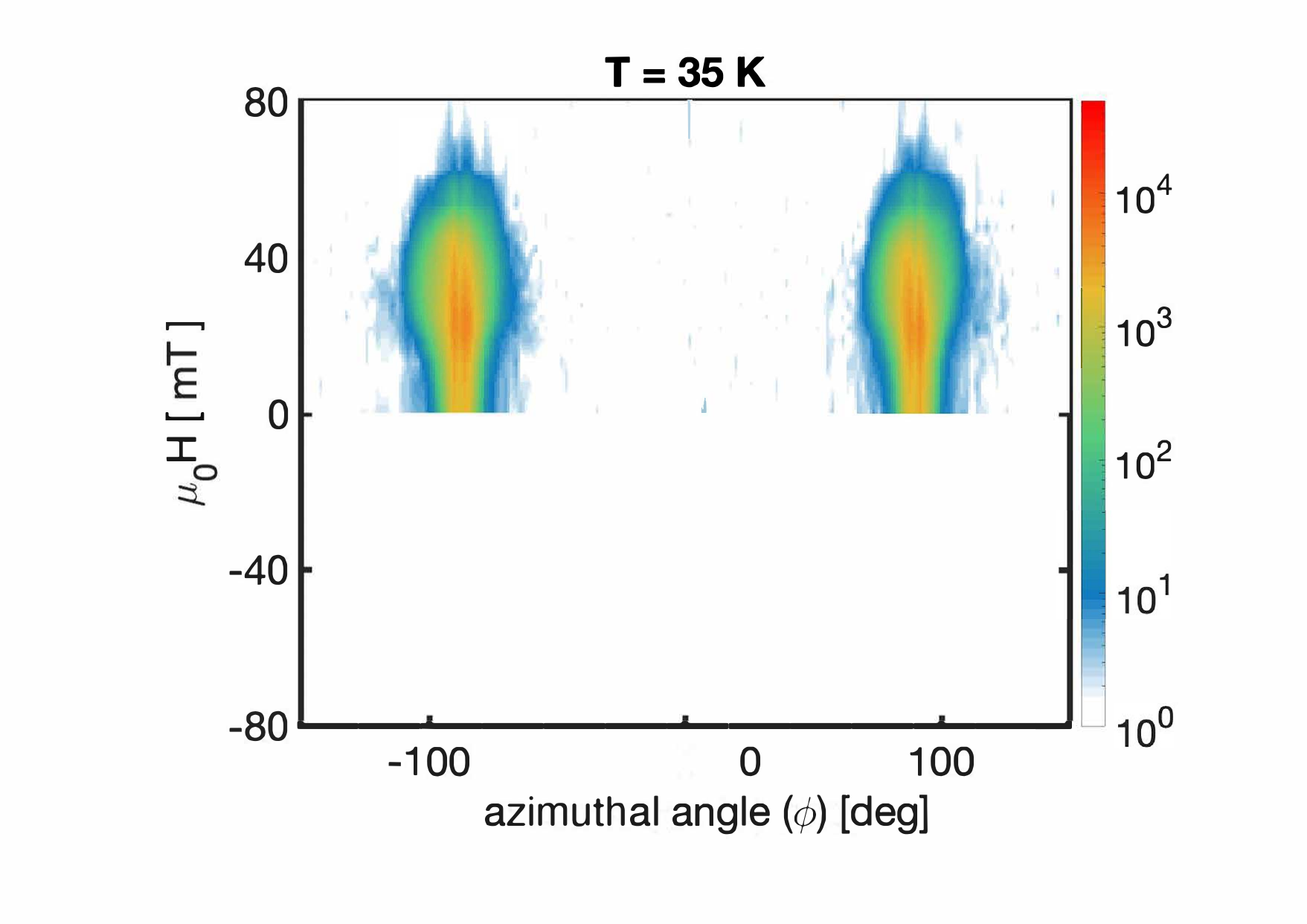}
    \caption{Contour plot of the scattered intensity recorded between  $Q$=0.05 and 0.013~nm$^{-1}$ as a function of  the azimuthal angle $\phi$ and the magnetic field for the T=35K. Here the measurements were performed only for positive fields. }
     \label{fig:35K}
\end{figure}
\end{minipage}
\vfill\null
\hfill

\begin{minipage}{2\linewidth}
\section{Simulations for $k_c = 0.07$ }

\begin{figure}[H]
    \centering
    \includegraphics[width = 0.99 \textwidth]{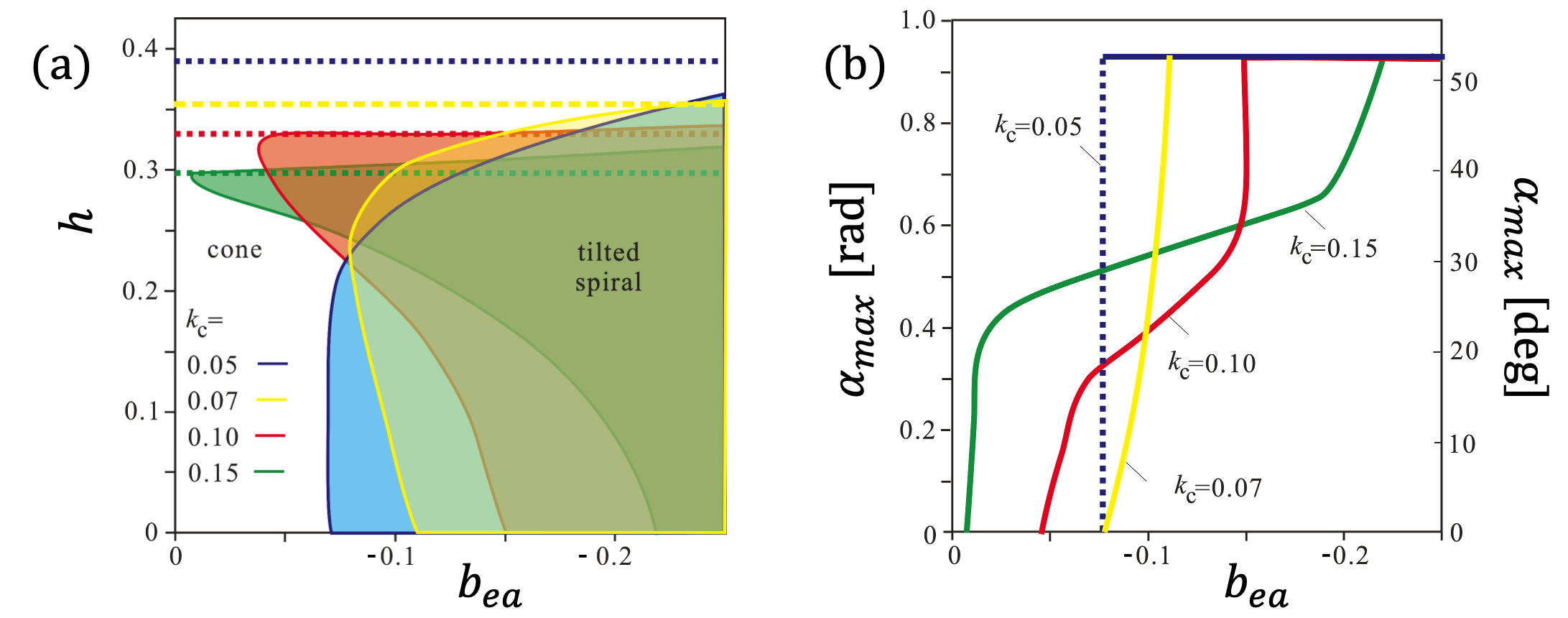}
    \caption{ Complementary results to Fig.7 revealing the behaviour for $k_c = 0.07$: (a)  phase diagram and (b)  maximum tilt angle plotted as a function of $b_{ea}$.}
     \label{fig:supp2}

\end{figure}
\end{minipage}

\vfill\null
\hfill

\pagebreak
\clearpage

\section{Energy minimization}

The non-linear partial differential Euler-Lagrange equations derived from the energy functional have been solved by numerical energy minimization procedure using finite-difference discretization on grids with adjustable grid spacings and periodic boundary conditions.
The components of the magnetization vector $\mathbf{m}$ have been evaluated in the knots of the grid, and for the calculation of the energy density we used finite-difference approximation of derivatives with different precision up to eight points as neighbors.
Furthermore, in order to check the stability of the numerical routines we additionally refined and coarsened the grids. 
For axial fields,  in order to reduce the artificial anisotropy incurred by the discretization, we used grid spacings $\Delta_y\approx\Delta_x$ leading to grids approximately square in the $xy$ plane.
The final equilibrium structure of the modulated states was obtained following an iterative procedure of the energy minimization using simulated annealing and a single-step Monte-Carlo dynamics with the Metropolis algorithm.
The details of the numerical methods used for the energy minimization are described in, e.g., Ref. \onlinecite{leonov2021} and hence will be omitted here.

In order to avoid an impediment introduced by the periodic boundary conditions, which necessarily arises due to the oblique spiral states when using a three-dimensional cube, we performed two-dimensional simulations.
For these, we wrote the anisotropy energy density  in a coordinate system connected with the wave vector of an oblique spiral and a corresponding plane of rotation. Thus, the axes of the new coordinate system $\acute{z}$ and $\acute{x}$ lie  in the plane $(1\overline{1}0)$, and the $\mathbf{Q}$-vector is aligned along $\acute{x}$ and at an angle $\alpha$ from the $z$-axis of an old coordinate system. 
Consequently, the energy density is minimized with respect to $\alpha$ leading to the optimum value of this angle characterising the oblique spiral state.

\end{appendices}

\end{document}